\documentclass[twocolumn,english,pra,showpacs]{revtex4}
\usepackage{amsfonts}
\usepackage{hyperref}
\usepackage{color}
\usepackage{dcolumn}
\usepackage{bm}
\usepackage{amssymb}
\usepackage[english]{babel}

\newcommand{\mf}[1]{\mathbf{#1}}
\newcommand{\bra}{\langle}
\newcommand{\ket}{\rangle}
\newcommand{\braket}[2]{\langle#1\vert #2\rangle}
\newcommand{\bracket}[3]{\langle#1\vert #2\vert #3\rangle}
\newcommand{\mt}{\mathrm}
\newcommand{\bs}[1]{\mbox{\boldmath${#1}$}}
\newcommand{\mc}{\mathcal}
\newcommand{\quintet}{$^5\Sigma_g^+$ }
\newcommand{\triplet}{$^3\Sigma_u^+$ }
\newcommand{\singlet}{$^1\Sigma_g^+$ }
\usepackage{graphicx}
\begin{document}

\title{Feshbach resonances in $^3\mathrm{He}^*- ^4\mathrm{He}^*$ mixtures}
\author{M.R.\ Goosen$^1$, T.G.\ Tiecke$^2$, W.\ Vassen$^3$, and
S.J.J.M.F.\ Kokkelmans$^1$}
\affiliation{$^1$Eindhoven University of Technology, P.O. Box 513, 5600MB Eindhoven, The
Netherlands}
\affiliation{$^2$Van der Waals-Zeeman Institute of the University of Amsterdam, 1018 XE
The Netherlands}
\affiliation{$^3$LaserLaB Vrije Universiteit, De Boelelaan 1081, 1081 HV Amsterdam,
The Netherlands}
\date{\today }

\begin{abstract}
\noindent We discuss the stability of homonuclear and heteronuclear mixtures of $^3$He and $^4$He atoms in the metastable 2 $^3$S$_1$ state (He*) and predict positions and widths of Feshbach resonances by using the Asymptotic Bound-state Model (ABM).
All calculations are performed without fit parameters, using \emph{ab-initio} calculations of molecular potentials. One promising very broad Feshbach resonance ($\Delta B=72.9^{+18.3}_{-19.3}~\mathrm{mT}$) is found that allows for tuning of the inter-isotope scattering length.
\end{abstract}

\maketitle

\section{Introduction}

\noindent The helium atom is one of the most simple atoms. Its electronic structure with only two electrons allows \emph{ab-initio} calculations of level energies with extreme precision, testing basic theory of atomic structure. Also for the interaction between two helium atoms \emph{ab-initio} quantum chemistry calculations allow highly accurate molecular potentials in some cases. In this paper we focus on helium atoms in the metastable 1s2s $^3$S$_1$ state (He*) for which molecular potentials have recently been calculated~\cite{muller:91, starck:94, dickinson:04, przybytek:05}. We investigate the possibilities to modify collision properties of mixtures of He* atoms in magnetic or optical dipole traps. Bose-Einstein Condensation (BEC) of $^4$He* atoms has been realized by several groups~\cite{robert:01, santos:01, tychkov:06, dall:07, ketterle:09}. This metastable isotope of helium has no nuclear spin, and is magnetically trappable in the fully stretched $|s,m_s\ket=|$1,+1$\ket$ state, where $s$ and $m_s$ are the electronic spin and magnetic quantum numbers respectively. The fermionic isotope $^3$He*, which, due to its nuclear spin $i$=1/2, shows hyperfine structure, has been cooled in the $f$=3/2 hyperfine manifold (with $\mf f = \mf s + \mf i$) to degeneracy by sympathetic cooling with $^4$He*~\cite{mcnamara:06}.

In the ultracold regime the scattering length $a$ accurately describes the interaction between the atoms. Magnetically tunable Feshbach resonances~\cite{chin:10} can be utilized to, in principle, tune the scattering length at will. Numerous experiments have been proposed based on this tuning possibility. For $^3$He*--$^4$He* mixtures it would be possible to observe phase separation when the scattering lengths are large and positive~\cite{molmer:98}. Using a position sensitive micro-channel plate detector to detect single atoms (with close to 100\% efficiency) may reveal boson-induced $p$-wave pairing of the $^3$He* fermions~\cite{efremov:02}. This detection technique has also allowed an atom-optics detection of the Hanbury Brown and Twiss effect for both bosons~\cite{schellekens:05} and fermions~\cite{jeltes:07}, and the possibility to tune interactions may also open up new research possibilities within this field~\cite{perrin:07}.

All experiments with ultracold He* atoms so far have been performed in fully stretched low-field seeking states ($|1,+1\ket$ for $^4$He*, $|f,m_f\ket=|$3/2,+3/2$\ket$ for $^3$He*). The atom-atom interactions for $^4$He* in the fully stretched spin states are known with astonishingly high precision~\cite{przybytek:05, przybytek:08}. From the binding energy of the $\nu=14$ ro-vibrational state in the quintet potential, which is only slightly below the dissociation threshold, a large positive scattering length $a$=7.567(24)~nm was deduced~\cite{mcnamara:06, przybytek:05, przybytek:08}, in very good agreement with experimental findings~\cite{moal:06}. For collisions between atoms which are not in the fully stretched spin states, Penning Ionization (PI) losses will strongly compromise the stability of trapped He* atoms in experiments. 

For the above mentioned reasons it has been favorable experimentally to prepare $^4$He* atoms in the fully stretched spin state. However, the absence of a nuclear spin in $^4$He* strongly limits the possibilities to tune the inter-atomic interactions via Feshbach resonances. Although it is possible to induce Feshbach resonances via the magnetic dipole interaction, a more efficient coupling to other states occurs when an internal hyperfine structure is present in at least one of the two interacting atoms. This occurs when we allow for mixtures between $^4$He* and $^3$He*. A recent experiment has already shown that $^4$He* atoms can be trapped in an optical dipole trap~\cite{partridge:10}, potentially allowing trapping of all magnetic substates of both helium isotopes. 

In this paper we discuss the possibility to access Feshbach resonances in collisions between $^4$He* and $^3$He* atoms. In addition, we also will consider homonuclear $^3$He*--$^3$He*  and $^4$He*--$^4$He* collisions. We will work with the Asymptotic Bound-state Model (ABM)~\cite{tiecke:10b} which will be reviewed in Sect.\ref{sect:ABM}. To determine the binding energies of the molecular potentials for the three possible isotope combinations, we use the accumulated phase method~ \cite{verhaar:09}. These binding energies will serve as input parameters for the ABM. In order to discuss experimentally relevant systems, we will evaluate the various possible loss mechanisms in Sect.\ref{sect:nonresonantloss} after which, possible Feshbach resonances will be explored in Sect.\ref{sect:FR}.

\section{The Asymptotic Bound-state Model}\label{sect:ABM}

To predict the magnetic field position and width of Feshbach resonances, we use the Asymptotic Bound-state Model (ABM) ~\cite{wille:08, tiecke:10a, tiecke:10b}. We recapitulate the basic principles of the ABM and show how this model can be applied to metastable helium. For an elaborate discussion on the ABM we refer the reader to Ref.~\cite{tiecke:10b}. In our approach, we ignore the effect of PI on the elastic interactions. A more detailed discussion of PI follows later in Sec.~\ref{sect:nonresonantloss}. To simplify the notation in discussing various helium isotope combinations, we use that $\mf{f}=\mf{s}$ in the absence of nuclear spin for $^4$He*, thereby allowing similar notational treatment of the spin structure of fermionic and bosonic helium atoms. 

\begin{figure}[!h]
\centering
\includegraphics[width=\columnwidth,keepaspectratio=true]{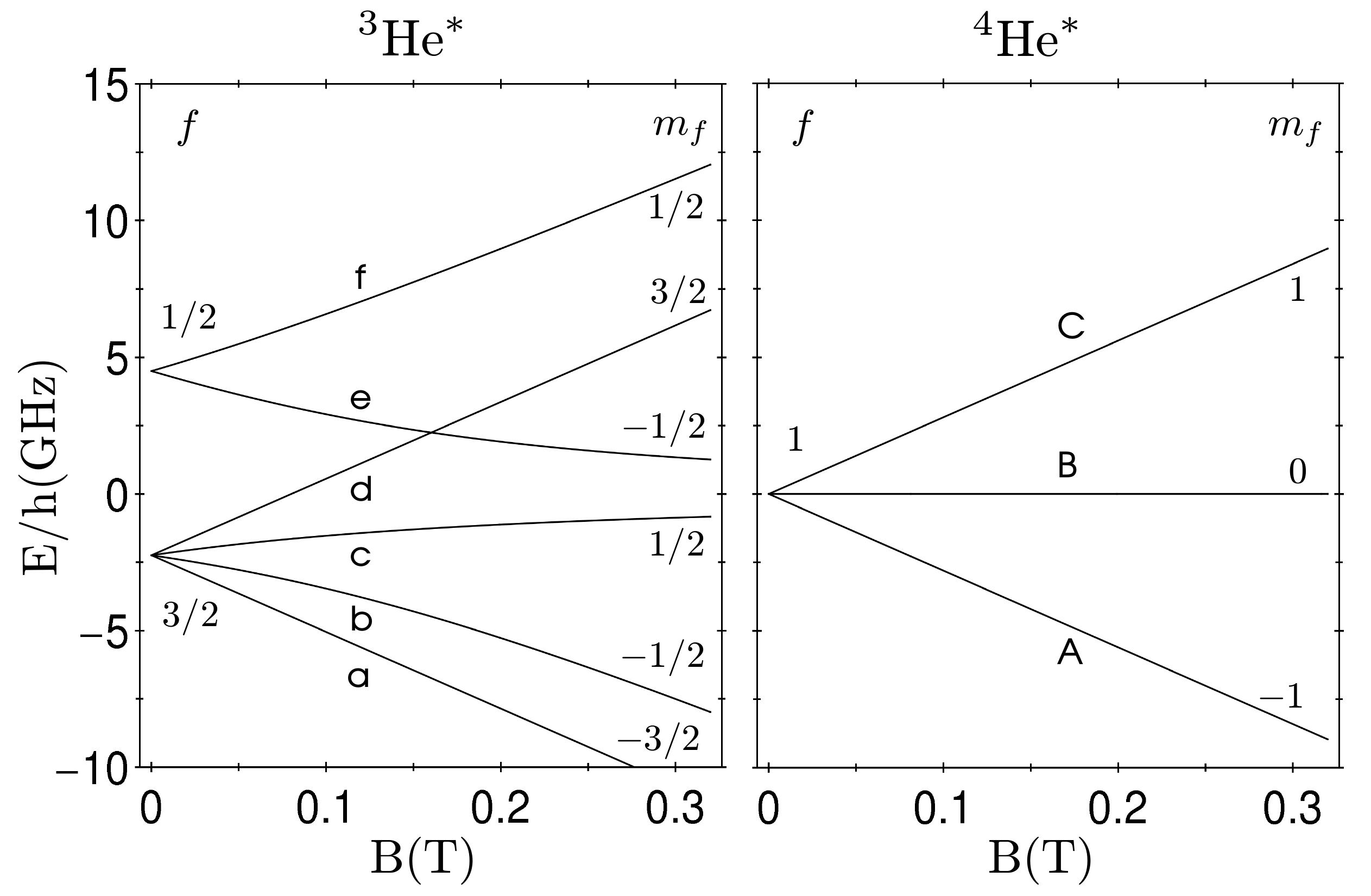}
\caption{Single-atom internal energy diagrams for $^3$He* and $^4$He*. The capital (lower case) letters are used to label the energies of $^4$He* ($^3$He*). The magnetic substates are also labeled by their $f,m_f$ quantum numbers, where at non-zero magnetic field $m_f$ is still a good quantum number, whereas $f$ is not. }\label{fig:1bodyinternal}
\end{figure}

The ABM enables us to determine the energy of the coupled molecular states, the eigenstates of the total Hamiltonian $\mc{H}$, without solving the actual coupled radial Schrodinger equation. For the collision of two metastable helium atoms in an external magnetic field the total Hamiltonian is
\begin{equation}
\mc{H}=\frac{\mf{p}^2}{2\mu}+\mc{H}_\mt{int}+\mc{V}+\mc{V}_\mt{dd},
\end{equation}
where the first term represents the relative kinetic energy with $\mu$ the reduced mass, and $\mc{H}_\mt{int}$ the two-body internal energy. The internal energy of $^3$He* and $^4$He* as a function external magnetic field is shown in Fig.~\ref{fig:1bodyinternal}. Note the inverted hyperfine structure of $^3$He*, which means that the $f=3/2$ manifold is below the $f=1/2$ manifold.

The central (Coulomb) interaction $\mc{V}$ depends on the magnitude of the total electron spin $\mf{S}=\mf{s}_1+\mf{s}_2$ and the distance between the nuclei $r=|\mf{r}|$. This interaction can be decomposed as
\begin{equation}
\mc{V}(r)=\sum_{S=0}^{2}\! ^{2S+1}V(r) \hat{P}_S,
\end{equation}
where $^{2S+1}V(r)$ is the adiabatic molecular potential for the molecular state $^{2S+1}\Sigma_{g,u}^+$, and $\hat{P}_S$ projects onto the $S$-subspace. For $^4$He*, short range molecular potentials have been calculated \emph{ab-initio} by M\"uller \emph{et al.}~\cite{muller:91}. In this paper, we refer to the molecular potentials as singlet $S=0$ (\singlet), triplet $S=1$ (\triplet) and quintet $S=2$ (\quintet) potentials. This nomenclature is based on the spin configuration of all four electrons of the two He* atoms. Note that often a single He* atom is also referred to as being in a spin-triplet state. Here, however, we do not use this designation to avoid confusion with the molecular singlet/triplet/quintet potentials.

The direct dipole-dipole interactions of the electronic spins~\footnote{We will not consider the much weaker electron-nuclear and nuclear-nuclear contributions to the magnetic dipole-dipole interaction.} $\mc{V}_{dd}$ can be written as a scalar product of two irreducible spherical tensors of rank 2:
\begin{equation}\label{eq:vdd}
\mc{V}_{dd}(\mf{r})=-\frac{3\alpha^2}{r^3}\sum_{q=-2}^{+2}(-1)^q\{\hat{\mf{r}}\otimes\hat{\mf{r}}\}_{2-q}\{\mf{s}_1\otimes\mf{s}_2\}_{2q},
\end{equation}
where $\{\hat{\mf{r}}\otimes\hat{\mf{r}}\}_{2m}=\sqrt{\frac{8\pi}{15}}Y_{2m}(\hat{r})$ is a spherical harmonic~\cite{AngularBook:88}, and  $\alpha$ is the fine structure constant. Being anisotropic, the interaction allows for redistribution of angular momentum between spin $\mf{S}$ and orbital $\bs{\ell}$ angular momentum. Therefore, although much weaker, this interaction can couple many more states as compared to isotropic interactions. The orbital angular momentum coupling can only occur for states when $\ell-\ell'=0,\pm2$, with the exception of $\ell=\ell'=0$ which is forbidden. The spin angular momentum coupling by $\mc{V}_{dd}$ obeys similar selection rules; $S-S'=0,\pm2$, where $S=S'=0$ is not allowed. 

From ultracold collisions it is well known that the position of the least bound state is crucial for the determination of the interaction properties. In the ABM we follow the same philosophy, and since the vibrational level splitting of the least bound levels for a light atom such as metastable helium is much larger (at least $h \times 21~\mathrm{GHz}$ for the $^4$He*--$^4$He* system, where $h$ is Planck's constant) than the hyperfine energy, we only use the properties of the least bound state. We expand the coupled solutions into a product basis of spin states and molecular states of the uncoupled $^{2S+1}V(r)$ potentials. These molecular states are the eigenstates of the relative Hamiltonian, defined as $\mc{H}_\mt{rel}\equiv \mf{p}^2/2\mu+\mc{V}$, and are denoted as $|\psi_{S,\ell}^\mc{A}\ket$ with corresponding binding energies $\epsilon_{S,\ell}^\mc{A}$. We have defined $\mc{A} \in \{ 33,34,44 \} $ to label the different isotope combinations which we consider. Having specified the Hamiltonian and the appropriate basis, the energies of the coupled bound states follow from a simple matrix diagonalization as a function of the strength of the magnetic field. 

To determine the characteristic properties of Feshbach resonances, i.e. their magnetic field widths and positions, we need to examine the behavior of the coupled bound states near the threshold of an open channel. A channel is called \emph{open} when the total energy exceeds the channel threshold energy, which is defined as the sum of the single atom internal energies. The intersection of the coupled bound state with this threshold gives the position of a Feshbach resonance, which is accurate to the order of the width of this resonance. However, in reality the coupled bound state will acquire an increasing open channel component as it approaches the open channel threshold. Therefore, near threshold the binding energy curve will bend quadratically towards the threshold curve as a function of magnetic field~\cite{chin:10}. From this behavior of the bound state it is possible to extract the width, and an improved estimate of the position of the resonance.

Since we distinguish between open and closed channels, it is useful to partition the space of states describing spatial and spin degrees of freedom into an open and a closed channel subspace~\cite{feshbach:58, feshbach:62}. The Hamiltonian for the system is written as
\begin{equation}
\mc{H}=\mc{H}_{PP}+\mc{H}_{QQ}+\mc{H}_{PQ}+\mc{H}_{QP},
\end{equation}
where $\mc{H}_{PP}\equiv P\mc{H}P,\mc{H}_{PQ}\equiv P\mc{H}Q$, etc. Here $P$ and $Q$ are projection operators onto the open and closed channel subspace respectively. The bound states of $\mc{H}_{QQ}$ will be responsible for  Feshbach resonances when coupled to the open channel subspace via $\mc{H}_{PQ}$($\mc{H}_{QP}^\dag$). These bare $\mathcal{Q}$-space bound states, denoted by $|\phi_Q\ket$, then become dressed by this interaction.  Usually this dressing occurs by coupling to scattering states in the open channel $\mathcal{P}$-space whereas for the ABM we use the bound states $|\phi_P\ket$ of the open channel subspace. The magnetic field at which the energy of the dressed bound state becomes degenerate with the energy of the threshold is where a Feshbach resonance will occur. 

The width of the resonance will depend on the coupling strength between the open and the resonant closed channels ($\mc{H}_{PQ}$), and the binding energy of the open- and closed-channel bound states ($\epsilon_P,\epsilon_Q$ respectively). We define the width $\Delta B$ as the difference in magnetic field where $a=0$ and $a=\infty$. For this purpose we introduce the $S$-matrix, which can be written for elastic scattering as $S=e^{2i\delta(k)}$ where $\delta(k)$ is the scattering phase, with $\hbar k=\sqrt{2\mu E}$, and the energy $E$ is the collision energy defined with respect to the open channel threshold energy. The scattering length is defined as $a\equiv-\lim_{k\to0} \tan\delta(k)/k$.

The energy of the dressed bound state, which is also referred to as the Feshbach molecular state, corresponds to a pole of the total $S$-matrix. This scattering matrix $S=S_PS_Q$ (of the effective problem in $\mathcal{P}$-space) is a product of the $S_P$ matrix, and a resonant part $S_Q$ which involves coupling of a $\mathcal{Q}$-space bound state to $\mathcal{P}$-space~\cite{moerdijk:95}. The $S_P$ scattering matrix, which describes the scattering process in $\mathcal{P}$-space in the absence of coupling to the $\mathcal{Q}$-space, 
is determined by considering only the dominant bound state in $\mathcal{P}$-space \cite{tiecke:10b}. Here we neglect the other nearby resonance poles in $\mathcal{P}$-space \cite{marcelis:04} which, is a valid approximation if the background scattering length ($a_\mt{bg}$) is larger than the typical range of the interaction potential \cite{tiecke:10b}. By determining the total $S$-matrix, we are able to deduce the (magnetic field) position and width of the Feshbach resonance. 

The $S_Q$ scattering matrix is usually determined by a single $\mathcal{Q}$-space bound state. However, it is possible that multiple $\mathcal{Q}$-space bound states have to be taken into account to properly describe a Feshbach resonance, as will be the case for the wide $^3$He*-$^4$He* resonance discussed later in Sect.~\ref{sect:FR}. 

The multiple $Q$-state expression for $S_Q$ is given by \cite{feshbach:62}
\begin{equation}\label{eqn:SQ}
S_Q=1-2\pi i\sum_n\frac{ \gamma^{(n)}}{E-E_n},
\end{equation}
where $E_n$ are the complex eigenvalues of $H_{QQ}+W_{QQ}$. The operator $W_{QQ}\equiv H_{QP}\frac{1}{E^+-H_{PP}}H_{PQ}$ describes a temporary transition from $\mathcal{Q}$-space to $\mathcal{P}$-space, propagation in $\mathcal{P}$-space, and re-emission into $\mathcal{Q}$-space \footnote{Here we define $E^+=E+i\delta$ with $\delta$ approaching zero from positive values}. These eigenvalues can be determined by solving the secular equation in the basis of $Q$-space bound states $|\phi_Q\ket$. By demanding unitarity of the $S$-matrix the coupling elements $\gamma^{(n)}$ can be expressed in terms of the  complex energies $E_n$ \cite{feshbach:62}. Hereby we have completely specified the total $S$-matrix. 

The free parameters in our model which determine the magnetic field resonance position $B_0$ and the field width $\Delta B$, are the binding energies $\epsilon_{S,\ell}^\mc{A}$ and the overlap between various molecular states $\braket{\psi_{S,\ell}^\mc{A}}{\psi_{S',\ell'}^\mc{A}}$~\cite{tiecke:10b}. Next we discuss how we can obtain these quantities by utilizing known molecular potentials. 

\section{Molecular states}

\label{sect:molecular} 
The essential parameters for the ABM are the binding energies $\epsilon_{S,\ell}^\mc{A}$ of the molecular potentials. If these values are known, Feshbach resonances can be predicted with an accuracy determined by the accuracy at which the $\epsilon_{S,\ell}^\mc{A}$ parameters are known. 
Vice-versa it is possible to obtain the binding energies by fitting the calculated resonance positions to experimentally observed resonances as has been shown in Ref. \cite{wille:08}. 
For metastable Helium no Feshbach resonances have been observed so far. To predict resonance positions we thus require knowledge of the binding energies of the $S=0, 1, 2$ potentials. These potentials are known from literature, however there is a significant difference between the $S=0$ and $S=1$ potentials on one side and the $S=2$ potential on the other side. The former ones are usually described as complex potentials to incorporate the effect of PI and have been described to a certain degree of accuracy by M\"uller, et al. \cite{muller:91}. The latter potential ($S=2$) does not exhibit PI and can therefore be accurately described by a purely real potential, which has been measured and calculated with very high accuracy \cite{przybytek:05,moal:06}.

In order to properly account for these uncertainties we make use of the \emph{accumulated phase method}~\cite{moerdijk:94,verhaar:09}. Moreover, since we study different isotope combinations, the uncertainties of the potentials are reliably treated via mass scaling within this method.
The $s$-wave accumulated phase can be written as
\begin{equation}\label{eqn:accumulated0}
^{\mc{A}}\phi_{S}(E)= {}^{\mc{A}}\phi_{S}^{(0)}(E)+\Delta^{\mc{A}}\phi_S,
\end{equation}
where $^{\mc{A}}\phi_{S}^{(0)}(E)$ is the uncorrected accumulated phase, resulting directly from the radial wavefunction in the inner region of a molecular potential $^{2S+1}V$, for a given energy $E$, at a radial range $r_0$. We allow for a phase correction $\Delta^{\mc{A}}\phi_S$, which accounts for the mismatch between calculated and experimentally measured quantities. It is determined from an asymptotic boundary condition, by demanding a particular energy for the highest bound state in the potential. The accumulated phase for a different isotope combination $\mc{A}'$ is found by a mass-scaling of the phase correction:
\begin{equation}\label{eqn:accumulated}
^{\mc{A}'}\phi_{S}(\epsilon_S^{\mc{A}'})= {}^{\mc{A}'}\phi_{S}^{(0)}(\epsilon_S^{\mc{A}'})+\mc{R}\Delta^{\mc{A}}\phi_S,
\end{equation}
where $\mc{R}=\sqrt{\mu_{\mc{A}'}/\mu_{\mc{A}}}$. We have verified the accumulated phase calculations by comparing with calculations performed on the full (complex) potentials. In the following we will discuss the different potentials in more detail.

\subsection*{S=2 potential}

A highly accurate \emph{ab-initio} $S=2$ potential was determined by Przybytek and Jeziorski~\cite{przybytek:05,przybytek:08}. For $^4$He* - $^4$He* they predicted a binding energy of the least bound state ($\nu=14$) equal to -89.6(8)~MHz. Using two-photon photo-association spectroscopy, Moal \emph{et al.}~\cite{moal:06} measured the binding energy of this least bound state to be $\epsilon^{44}_2/h=-91.35(6)$~MHz. We use this measurement as input parameter for $^4$He* - $^4$He* ABM calculations, and we use Eq.~(\ref{eqn:accumulated}) to find the binding energies for the other isotope combinations.

We construct the full $S=2$ potential by fitting the short-range potential by Przybytek~\cite{przybytek:05} to the long-range dispersive potential $V_\mt{disp}(r)=-C_6/r^6-C_8/r^8-C_{10}/r^{10}$ using the accurately known dispersive coefficients~\cite{yan:98}. We forge these potentials around $r=20\,\mathrm{a_0}$ by vertically shifting the short-range potential to match the long-range potential and by applying a smoothing function. By the demand of a bound state at $\epsilon^{44}_2/h=-91.35(6)$~MHz, we apply the phase correction at $r_0=18\,\mathrm{a_0}$, which is determined with Eq.~(\ref{eqn:accumulated0}). The results for the different isotope combinations are $\epsilon_{2}^{34}/h=-4.84$~MHz and $\epsilon_{2}^{33}/h=-413.83$~MHz. These energies compare well to those obtained by Przybytek and Jeziorski~\cite{przybytek:tobepublished} based on the full scattering potentials.

\begin{figure}[!h]
\centering
\includegraphics[width=\columnwidth,keepaspectratio=true]{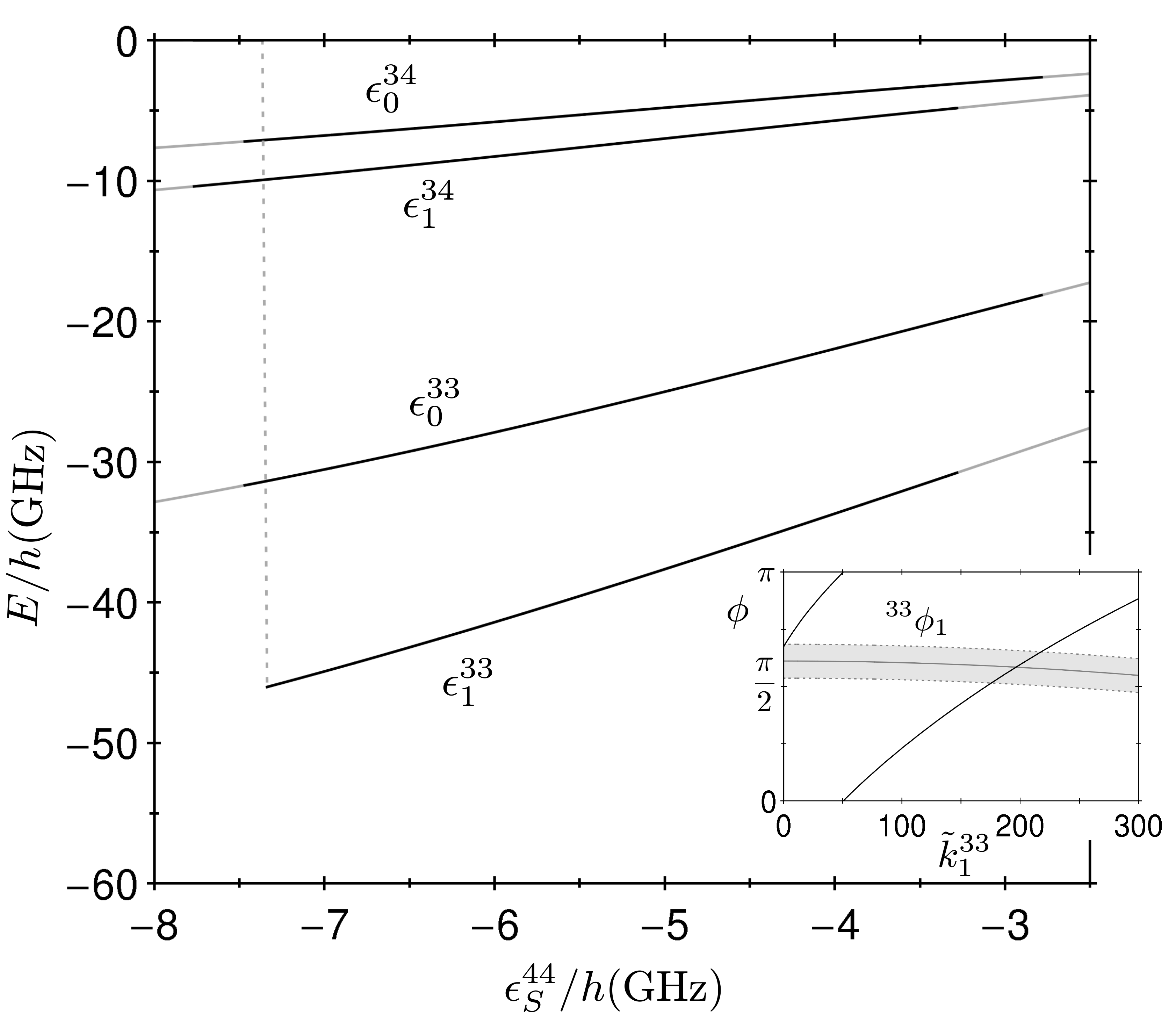}
\caption{Variation in the binding energies of the \emph{least} bound state of the $S=0,1$ potentials. The uncertainty in $\epsilon_{S}^{44}$ leads to a range of mass-scaled binding energies, indicated by thick black lines. Within the present accuracy, the triplet $^3$He* - $^3$He* potential can become deep enough to capture a new bound state. This weakly bound state cannot be distinguished on this scale. The inset demonstrates how we find the binding energies for this case. The accumulated phase of Eq.~(\ref{eqn:accumulated0}) (which is indicated by a grey shaded area) is matched with the phase of the wavefunction (solid line), starting from an asymptotic bound state boundary condition. These accumulated phases are plotted as a function of $\tilde{k}_1^{33}\equiv\sqrt{|\epsilon_1^{33}/h|}~(\mathrm{MHz}^{1/2})$. The variation in the accumulated phase, caused by the uncertainties of the potential, corresponds to a range of binding energies. Therefore, $\tilde{k}_1^{33}$ can take any of the values $0 < \tilde{k}_1^{33}< 1~\mathrm{MHz}^{1/2}$ or $175 < \tilde{k}_1^{33} < 215~\mathrm{MHz}^{1/2}$}\label{fig:bindingenergy}
\end{figure}

\subsection*{S=0,1 potentials}

The short range parts of the $S=0,1$ potentials, as obtained by M\"{u}ller \emph{et al.}~\cite{muller:91}, are known accurately enough to calculate the position of the least bound levels in these potentials with relatively large errorbars. The $S=0,1$ potentials also include an imaginary part to incorporate for the PI losses. As the ABM works with only real binding energies we assume real $S=0,1$ potentials. This introduces an additional error to which we will come back later. The use of the accumulated phase method allows us to incorporate these different uncertainties in the $\Delta^{\mc{A}}\phi_S$ parameter. 

We use results obtained by Leo \emph{et al.}~\cite{leo:01} to set upper and lower bounds on the scattering lengths for these potentials
\begin{equation}\label{eq:leoScat}
a_0^{44}=34(10) a_0,\qquad  a_1^{44}=32(9) a_0,
\end{equation}
where the error bars indicate the maximum inaccuracy of both the real and complex short range potentials as found by Leo \emph{et al.}~\cite{leo:01}. The scattering lengths enable us to determine binding energies for the bosonic metastable helium system, as well as for the other isotope combinations. The results of these calculations are summarized in Fig.~\ref{fig:bindingenergy}. 

As mentioned before scattering in the $S=0,1$ potentials experience strong PI losses which are generally described by including an imaginary term in the potential. Up till now we have neglected this imaginary terms, however, to calculate the binding energies we use the scattering lengths (Eq.~(\ref{eq:leoScat})) which are based on the full optical potentials. Therefore, this procedure accounts partly for the inaccuracy in the binding energies induced by using a real scattering potential.

Additionally, we have verified that using only real potentials induces a relatively small error by comparing the value of $\epsilon^{34}_1$ obtained with the accumulated phase method with the value obtained using the complete $S=1$ potential. For the imaginary potential we use the autoionization width as given in Ref. \cite{muller:91}. The total $S=0,1$ potentials were constructed by vertically shifting the short range potentials~\cite{muller:91} to (smoothly) match $^5V(r)-V_\mt{exch}(r)$ for $r>11.5\,\mathrm{a_0}$. The exchange terms, which depend on quantum number $S$, were determined in Ref.~\cite{tang:98}.  We find that including the autoionization width shifts the real part of $\epsilon^{34}_S$ by $\simeq h\times 1.5$~GHz. This is a significant shift, but still smaller than the uncertaincies in the binding energies caused by the errors as given in Eq.~ (\ref{eq:leoScat}). Therefore, the limiting factor in the present calculations is the inaccurately known short range potentials rather than the approximation of using only real potentials.

For the $S=0,1$ potentials of $^3$He*--$^3$He* we find that the scattering lengths are negative; i.e. virtual bound states will dominate the low energy scattering properties of these potentials~\cite{marcelis:04}. Within the variation of the $S=1$ potential, it is even possible that the virtual bound state turns into a (weakly) bound state, and therefore becomes the highest bound state of the potential which will make the scattering length (large) positive. This is displayed in the inset of Fig.~\ref{fig:bindingenergy}. The low energy scattering properties of the $S=1$ potential will depend strongly on the precise position of such a weakly bound state, which will inhibit accurate predictions for Feshbach resonances without more accurate knowledge of the $S=1$ potential.

\section{Inelastic decay processes}

\label{sect:nonresonantloss} To determine which two-body hyperfine states are best suited to explore Feshbach resonances, we discuss various loss mechanisms that may occur. A particular combination of states can be relatively stable, however, the occurrence of Feshbach resonances may limit the stability, since the underlying resonant bound state can suffer strongly from inelastic effects. We discuss the possibility of resonantly enhanced losses in the next section.

In contrast to ground-state atoms, metastable helium atoms can undergo highly exothermic ionizing collisions, since the internal energy of two He* atoms exceeds the He* ionization potential by 15 eV. These reactions, described by 
\begin{equation}
 \mt{He}^*+\mt{He}^* \rightarrow \bigg\{
\begin{array}{l}
     \mt{He}+\mt{He}^++\mt{e}^-  \\ 
     \mt{He}_2^++\mt{e}^-, 
 \end{array}
\end{equation}
will be referred to as Penning Ionization (PI), which includes the process commonly referred to as associative ionization. As this reaction is electrostatic, the total electron spin $\mf{S}$ is conserved (Wigner spin-conservation rule). For fully stretched states, where $S=2$, the reaction would violate spin conservation and is therefore forbidden in first order, whereas when $S=0$ or $S=1$ the reaction can proceed. The probability for PI for these latter potentials is $\sim$0.975~\cite{muller:91}, hence a severe loss process for collisions involving scattering through the singlet or triplet potential.

\begin{figure}[!h]
\centering
\includegraphics[width=\columnwidth,keepaspectratio=true]{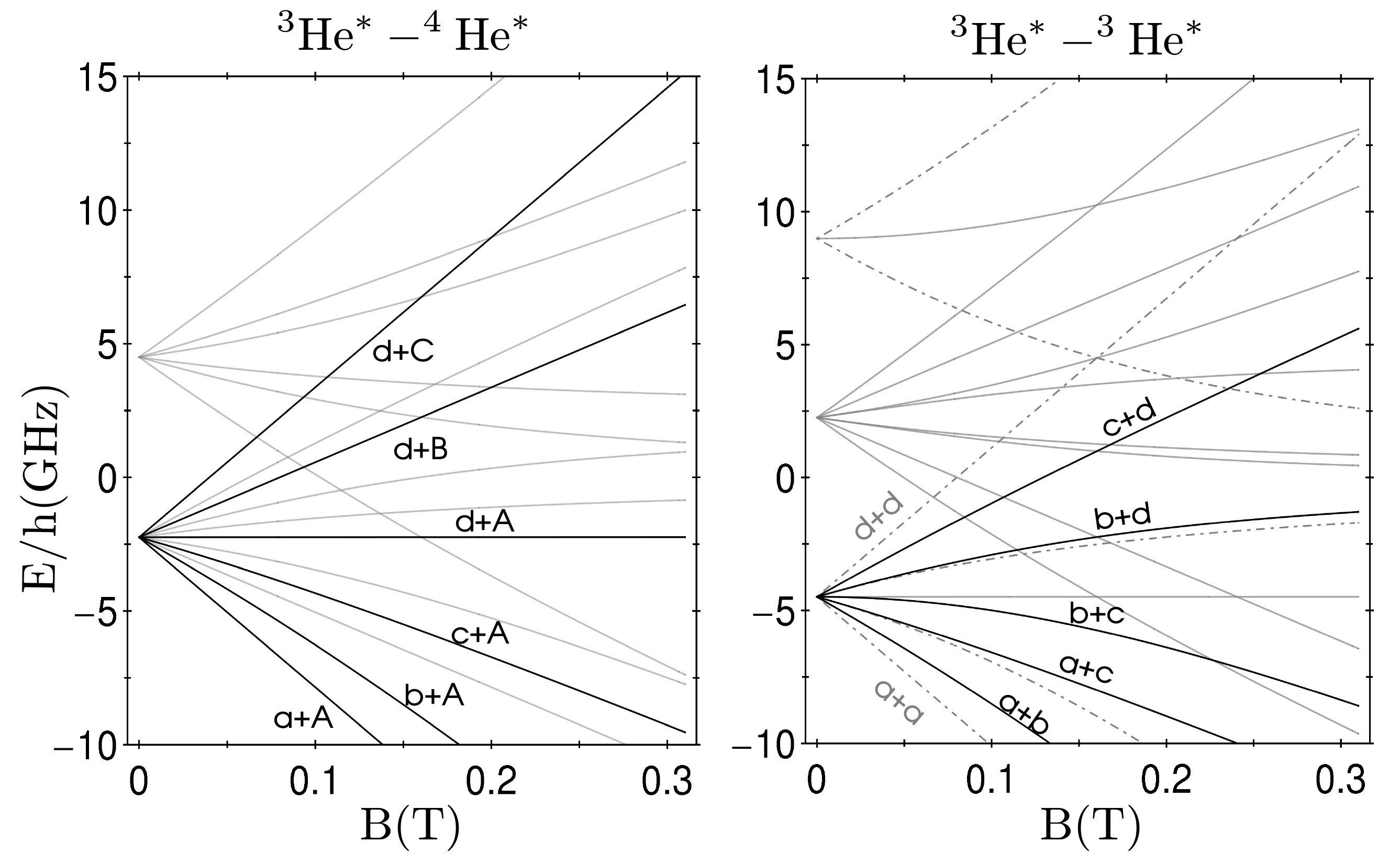}
\caption{The two-body hyperfine diagrams are shown for $^3$He*--$^4$He* and $^3$He*--$^3$He*. For each possible value of $m_F$, the energetically lowest two-body hyperfine state (i.e.~stable against spin exchange relaxation) is labeled (black), whereas the other states (unstable against spin exchange relaxation) are not labeled (gray). For homonuclear fermionic helium, symmetry prevents that hyperfine states composed of identical substates (dot-dashed) can be populated with $\ell=\mt{even}$.}\label{fig:2bodyinternal}
\end{figure}

Another important loss mechanism is spin exchange relaxation which is induced by the central part $\mc V$ of the interaction. For these isotropic interactions the projection $m_F=m_{f_1}+m_{f_2}$ of total spin angular momentum $\mf{F}=\mf{f}_1+\mf{f}_2$ on the magnetic field axis is conserved during the collision. By preparing atoms in the energetically lowest two-body hyperfine state, for a particular value of $m_F$, only unfavorable endothermic collisions can occur, effectively suppressing spin exchange relaxation losses. In Fig.~\ref{fig:2bodyinternal} we have labeled these two body states by their one-body constituents. 

The spin-dipole interaction $\mc V_{dd}$ between the spins of both electrons induces losses as well. For these anisotropic interactions the projection of total angular momentum $\bs{\mc{F}}=\mf{F}+\bs{\ell}$ is conserved, coupling only two-body states of the same $m_\mc{F}=m_F+m_\ell$. 
Here, two mechanisms may cause trap loss: spin relaxation ($\alpha_{rel}$) and relaxation-induced ionization ($\alpha_{ri}$). The first process only relaxes the spin projection $m_S$, whereas for the second process the $S=2$ state is coupled to $S=0$, which decays via the 'normal' PI mechanism. At low temperatures and low $B$ fields the latter process dominates. 
These mechanisms were found to be the most prominent cause of losses in a spin-polarized gas of $^4$He*. The loss rate was calculated to be four orders of magnitude smaller~\cite{fedichev:96}, as compared to an unpolarized gas, where direct PI is dominant for trap loss.

Generically three-body loss rates depend on the value of the two-body scattering length. For a homonuclear gas, where 3 (or 2) identical bosons or 2 identical fermions participate in a three-body collision, the three body loss rate will vary as $|a|^p$, where $p>3$~\cite{incao:05}. Therefore large values of $|a|$ are predicted to (strongly) enhance three-body loss rates, which are affected by spin state and statistics of the participating atoms in the process. 

The dominant loss mechanisms for trapped metastable helium atoms are Penning ionization and spin exchange relaxation. Therefore we will only consider two-body hyperfine states which are stable against these decay processes. Since symmetry will impose additional constraints, we will elaborate on the loss mechanisms for the homonuclear and heteronuclear case separately, in order to find the right experimental conditions (a sufficiently long lifetime) to search for Feshbach resonances. Future experiments based on predictions of the ABM discussed in Sect.~\ref{sect:ABM} will depend heavily on these background losses.

\subsection*{Homonuclear losses}

For homonuclear collisions (between identical bosons or fermions), the symmetrization requirement
\begin{equation}\label{eq:sym}
S+I+\ell=\mt{even},
\end{equation}
for their two-body state can have severe consequences on the stability of the gas. If we for example consider collisions of $^4$He* atoms, Penning ionization losses via the $S=1$ potential can only occur via $\ell=$odd collisions. These losses are therefore strongly suppressed at $T\sim1\mu$K, where the experiments around degeneracy take place~\cite{venturi:00}. 

For bosonic helium atoms all two-body hyperfine states except the degenerate $\mt{B+B,A+C}$ states are stable against spin exchange relaxation. Additionally, PI will also induce losses for these two states making them unsuitable in our pursue of Feshbach resonances. The fully stretched states $\mt{A+A,C+C}$ however, are stable against PI losses. If we restrict the temperatures such that we can consider only $s$-wave collisions, the $\mt{A+B,B+C}$ states will also only scatter via the $S=2$ potential. The stability of this mixture will however be limited by the losses between atoms in the $\mt{B}$ state. The stability can be improved by making $\mt B$ the minority spin species. In a magneto optical trap PI losses have been studied in the presence of MOT light and without. Good agreement between theory and experiment was obtained for loss rates of unpolarized atoms in the dark; the two-body loss rate turned out to be very large: $K_{44}^{\text{(unpol)}}$=1$\times$10$^{-10}$ cm$^3$/s~\cite{stas:06}. 

For fermionic helium atoms, ultracold collisions between atoms in the same substates can only occur via odd partial waves (Pauli principle), effectively stabilizing the gas against PI losses at $\mu$K temperatures. For $^3$He*, studies of losses in a MOT have also been performed and also here theory and experiment agree on the loss rate in the dark: $K_{33}^{\text{(unpol)}}$=2$\times$10$^{-10}$ cm$^3$/s~\cite{stas:06}. In the dark a mixture of all four magnetic substates of the trapped $f$=3/2 hyperfine manifold shows a loss rate that is even larger than for $^4$He*. The loss rate here is described very well assuming only ionizing collisions between atoms in different magnetic substates. 

\begin{figure}[!h]
\centering
\includegraphics[width=\columnwidth,keepaspectratio=true]{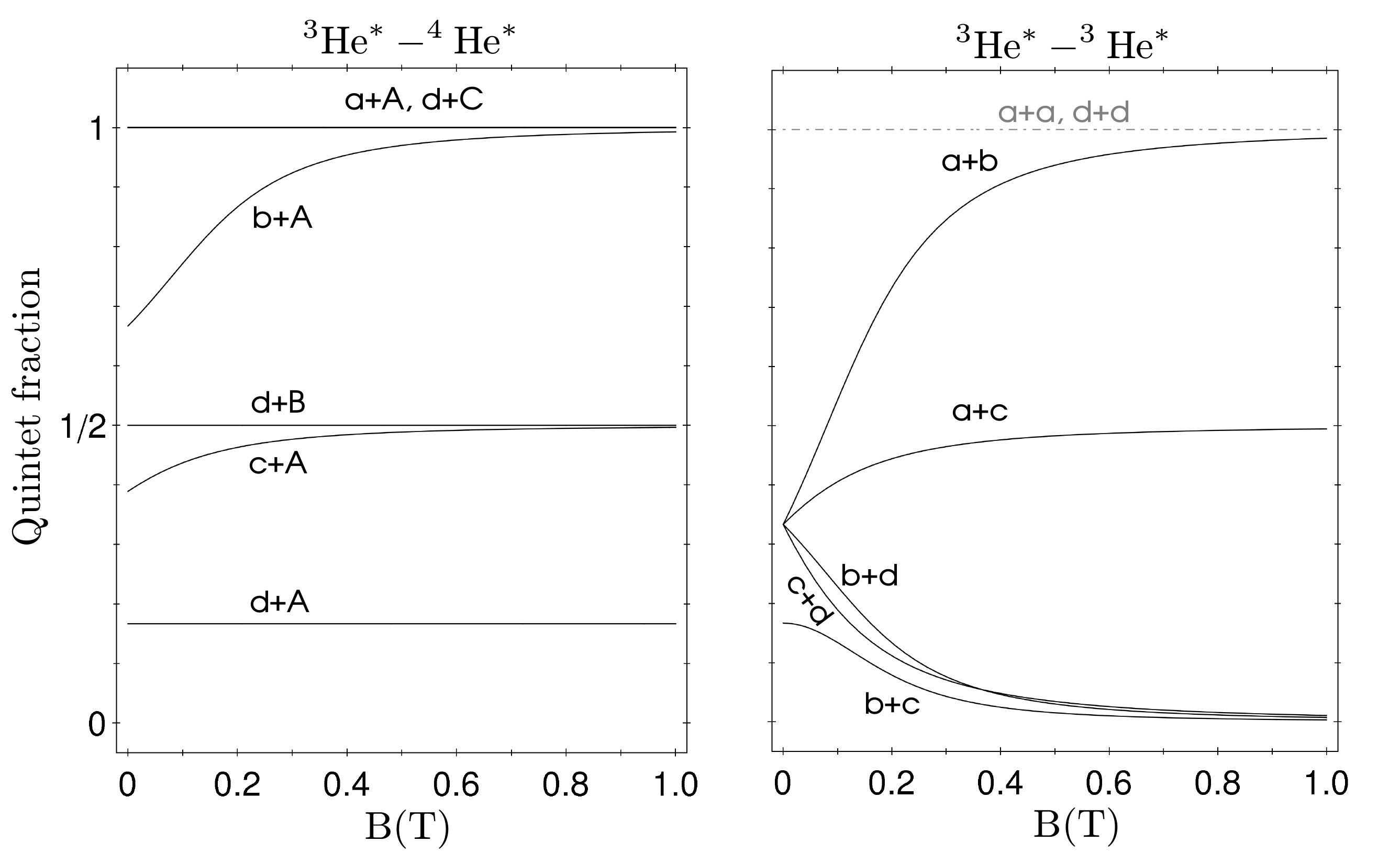}
\caption{The quintet fraction for two-body hyperfine states as a function of magnetic field is shown for $^3$He*--$^4$He* and $^3$He*--$^3$He*. We only show states which are stable against spin exchange relaxation, i.e. the energetically lowest two-body state for a particular $m_F$ value. The $\mt{a+a}$ and $\mt{d+d}$ states (gray) can only have collisions with $\ell=$odd due to symmetry. }\label{fig:qfrac}
\end{figure}

The nuclear spin of $^3$He* gives rise to magnetic field dependent PI loss. Therefore, we have plotted the $S=2$ fraction of the two-body hyperfine states (stable against spin exchange relaxation) as a function of the magnetic field in Fig.~\ref{fig:qfrac}. This decomposition into molecular states containing quantum number $S$ of the two-body hyperfine state allows us to estimate the Penning ionization rate which will be large for states containing a small $S=2$ fraction~\cite{stas:06}. 
For $s$-wave collisions of two fermions, the $\mt{a}+\mt{b}$ state ($m_F=-2$) will scatter via dominantly the $S=2$ potential at higher magnetic fields whereas for all the other states unfavorable PI processes will dominate. If we consider $\ell=$odd collisions as well, the two-body states $\mt{a+a,d+d}$ are stable against PI losses.

In the experiments on degenerate gases of either $^3$He* or $^4$He*, three-body losses have been shown to be less important than two-body losses. In the case of the fully stretched $\mt{C+C}$ state a three-body loss rate constant $L$=2$\times$10$^{-27}$ cm$^6$/s$^2$ was calculated~\cite{fedichev:96}, which experimentally was confirmed in BEC lifetime studies~\cite{tychkov:06} for $^4$He*.

\subsection*{Heteronuclear losses}

For collisions between $^3$He* and $^4$He* atoms we do not have symmetrization requirements as for the homonuclear case. Only for atoms in the fully stretched states PI is suppressed~\cite{stas:06, mcnamara:07}. In a MOT containing both isotopes a heteronuclear ionization rate coefficient was deduced from loss measurements that also agrees with theory: $K_{34}^{\text{(unpol)}}$=3$\times$10$^{-10}$ cm$^3$/s~\cite{mcnamara:07}. This shows that losses in an unpolarized He* gas at mK temperatures, for all isotopes, are well understood and may be extrapolated to $\mu$K temperatures. 

Considering heteronuclear collisions, we have to take into account that only $^3$He* and $^4$He* substates should be selected which are intrinsically stable. The previous discussion on the homonuclear situation then dictates that $^4$He* atoms in the $\mt{B}$ state have to be excluded. The decomposition into the quintet fraction of the possible entrance states as a function of magnetic field is shown in Fig.~\ref{fig:qfrac}. We find that in addition to the fully stretched $\mt{a+A,d+C}$ states, the $\mt{b+A}$ state will contain a large quintet fraction.

As the quintet scattering length is extremely large, we expected that the stability of a boson-fermion mixture of spin-polarized $^3$He* and $^4$He* atoms is severely compromised by three-body recombination processes. Initial observations indeed confirm this~\cite{mcnamara:06}. Recent experiments on a boson-fermion mixture of $^{87}$Rb--$^{40}$K Simoni, \emph{et al.}~\cite{simoni:08} however, have shown that such a system can be made more stable by having an excess of the fermions rather than the bosons. This is explained by the fact that for a three-body loss process in that case two identical fermions have to come close to each other which is suppressed by the Pauli principle. 

\subsection*{Effect on Feshbach resonances}

In view of inelastic loss processes, the selected stable two-body hyperfine states for all three helium isotope combinations are: $^4$He*--$^4$He*: $\{\mt{A+A,C+C}\}$, $^3$He*--$^3$He*: $\{\mt{a+a,a+b,d+d}\}$, and $^3$He*--$^4$He*: $\{\mt{a+A,b+A,d+C}\}$. Before we discuss the possibility of finding Feshbach resonances for these two-body states, we will consider the impact that inelastic loss processes might have on these resonances.

The presence of inelastic loss processes such as spin relaxation, relaxation-induced ionization, and Penning ionization
will affect the Feshbach resonances. These inelastic events may occur not only in the open $\mc{P}$ but also in the closed $\mc{Q}$ channel subspace. The scattering length describing the two-body interactions, including these effects, will now be complex valued~\cite{kempen:02}, and the divergence of the real part of the scattering length at resonance will be suppressed. The strength of the resonance will depend on the relative magnitudes of the coupling elements between the resonant state to the elastic and inelastic channels~\cite{hutson:07}.

Although Feshbach resonances are usually associated with various enhanced (two and three body) loss processes, they can also have a stabilizing effect \cite{smirne07, hutson:09}. Since inelastic losses can be induced in both the $\mc{P}$- and $\mc{Q}$-space by PI, it is possible that PI losses can be suppressed as the admixture of $\mc{P}$- and $\mc{Q}$-space can change in the vicinity of a Feshbach resonance.

For PI losses, two metastable helium atoms (in close proximity of each other, for $S\ne2$) couple to energetically lower ionic states which yields an inelastic process. To describe the effect of PI losses, one usually takes optical potentials and avoids the use of the ionic channels. This transforms the closed quantum system to an open one, i.e. the effective Hamiltonian describing the collision between two helium atoms has become non-Hermitian. This is different as compared to the inelastic scattering caused by the magnetic dipole-dipole interaction, which is described by increasing the number of channels in the open channel subspace but keeping the effective Hamiltonian Hermitian. Although the description of both processes is different, the effect of these processes will (in principle) be the same.

The PI process will also influence the molecular states of the $S=0,1$ potentials. The generic effect of this imaginary potential to describe PI, is that it will cause the bound states of the real potential to become unstable~\cite{dabrowski:96}, i.e. they acquire a finite lifetime. The binding energies of such unstable bound states are complex valued, as one would expect for a non-Hermitian Hamiltonian. The imaginary part of the complex energy of such an unstable bound state is not used as a parameter in the ABM.

\section{Feshbach resonances}\label{sect:FR}

Considering the various inelastic decay mechanisms due to PI and spin exchange relaxation, we have narrowed down the number of interesting open (also referred to as entrance) channels dramatically. In our search for Feshbach resonances our focus will be on resonances caused by coupled bound states which are mainly in an $S=2$ state. The reason for this lies in the inaccuracy of the $S=0,1$ potentials which will lead to a significant spread of possible singlet and triplet binding energies~\footnote{Although there is a spread in the possible binding energies we will use the nominal values for all calculations.}, as has been discussed in Sect.~\ref{sect:molecular}. We limit the search for Feshbach resonances to magnetic fields up to $1$~T. At the end of this section, the found Feshbach resonances are summarized in Table~\ref{tab:FRsummary}. 

Feshbach resonances induced by the magnetic dipole-dipole interaction $\mc{V}_{dd}$ will also be considered, although these are expected to yield much weaker Feshbach resonances as compared to the ones induced by central interactions $\mc{V}$. Since the  dipole-dipole interaction is weak we consider only first order processes and thus only include partial waves $\ell\le2$. 
The basis set for the ABM calculations will consist of the (bound) eigenstates $|\psi_{S,\ell}^\mc{A}\ket$ and energies $\epsilon_{S,\ell}^\mc{A}$ of the $^{2S+1}V(r)+\ell(\ell+1)\hbar^2/(2\mu r^2)$ potentials for $\ell\le2$. These molecular states are determined in a similar fashion as presented in Sect.\ref{sect:molecular} for $s$-wave bound states. The eigenstates $|\psi_{S,\ell}^\mc{A}\ket$ allow us to determine the matrix elements of $\mc{V}_{dd}$ as 
\begin{eqnarray}
\bra\Psi_{S,\ell}^\mc{A}|\mc{V}_{dd} |\Psi_{S',\ell'}^\mc{A}\ket=-3\alpha^2
\bracket{\psi_{S,\ell}^\mc{A}}{\frac{1}{r^3}}{\psi_{S',\ell'}^\mc{A}}\sum_{q=-2}^2 (-1)^q 
\nonumber\\ \sqrt{\frac{8\pi}{15}}
\bracket{\ell m_\ell}{Y_{2-q}(\hat{r})}{\ell' m_{\ell'}} 
\bracket{\sigma}{\{\mf{s}_1\otimes\mf{s}_2\}_{2q}}{\sigma'},
\end{eqnarray}
where $|\Psi_{S',\ell'}^\mc{A}\ket\equiv|\psi_{S,\ell}^\mc{A}\ket|\ell m_\ell\ket|\sigma\ket$, and $|\sigma\ket$ is the two-body spin state~\cite{tiecke:10b}. The $r$-dependent factor $\bracket{\psi_{S,\ell}^\mc{A}}{r^{-3}}{\psi_{S',\ell'}^\mc{A}}$ is determined by a numerical integration. By diagonalizing the Hamiltonian $\mc{H}$ we find the energies of the coupled bound states, as described in Ref.~\cite{tiecke:10b}. The magnetic dipole-dipole interaction is thus treated as a perturbation in first order in the ABM, similar to the treatment of magnetic dipole-dipole interactions for metastable helium atoms by Beams \emph{et al.}~\cite{beams:06}. 

We selected our entrance channels by minimizing possible inelastic losses due to PI and spin exchange relaxation, however, $\mc{V}_{dd}$ may couple to other states which are significantly less stable than these entrance channels. In view of the Feshbach formalism discussed in Sect.~\ref{sect:ABM}, these inelastic processes occur when the dimension of the open channel subspace $\dim(\mc{P})>1$. The inelastic losses induced by $\mc{V}_{dd}$ will alter the Feshbach resonance characteristics, e.g. the real part of the scattering length will not diverge on resonance as it would if there was only elastic scattering~\cite{hutson:07}. Throughout this section, just as with bare closed channel bound states, we only use the dominant (i.e. energetically closest to the threshold) open channel bound state for the determination of $B_0,\Delta B$. We neglect the effect of the inelastic losses due to $\mc{V}_{dd}$ and PI on the Feshbach resonances.

\subsection*{Homonuclear gas}

For $^4$He*--$^4$He* collisions, the absence of nuclear spin prohibits (coupled) bound states within an $m_F$-manifold to cross the open channel threshold and induce Feshbach resonances. Since $\mc{V}_{dd}$ can couple states of different $m_F$, Feshbach resonances can be induced by the spin-spin interaction. For the two-body state $\mt{A+A}$ we find two Feshbach resonances. It is important to note that the entrance channel $\mt{A+A}$ is purely quintet and there is a bound state $\epsilon^{44}_2/h=-91$~MHz parallel to the entrance channel. 

\begin{figure}[!h]
\centering
\includegraphics[width=\columnwidth,keepaspectratio=true]{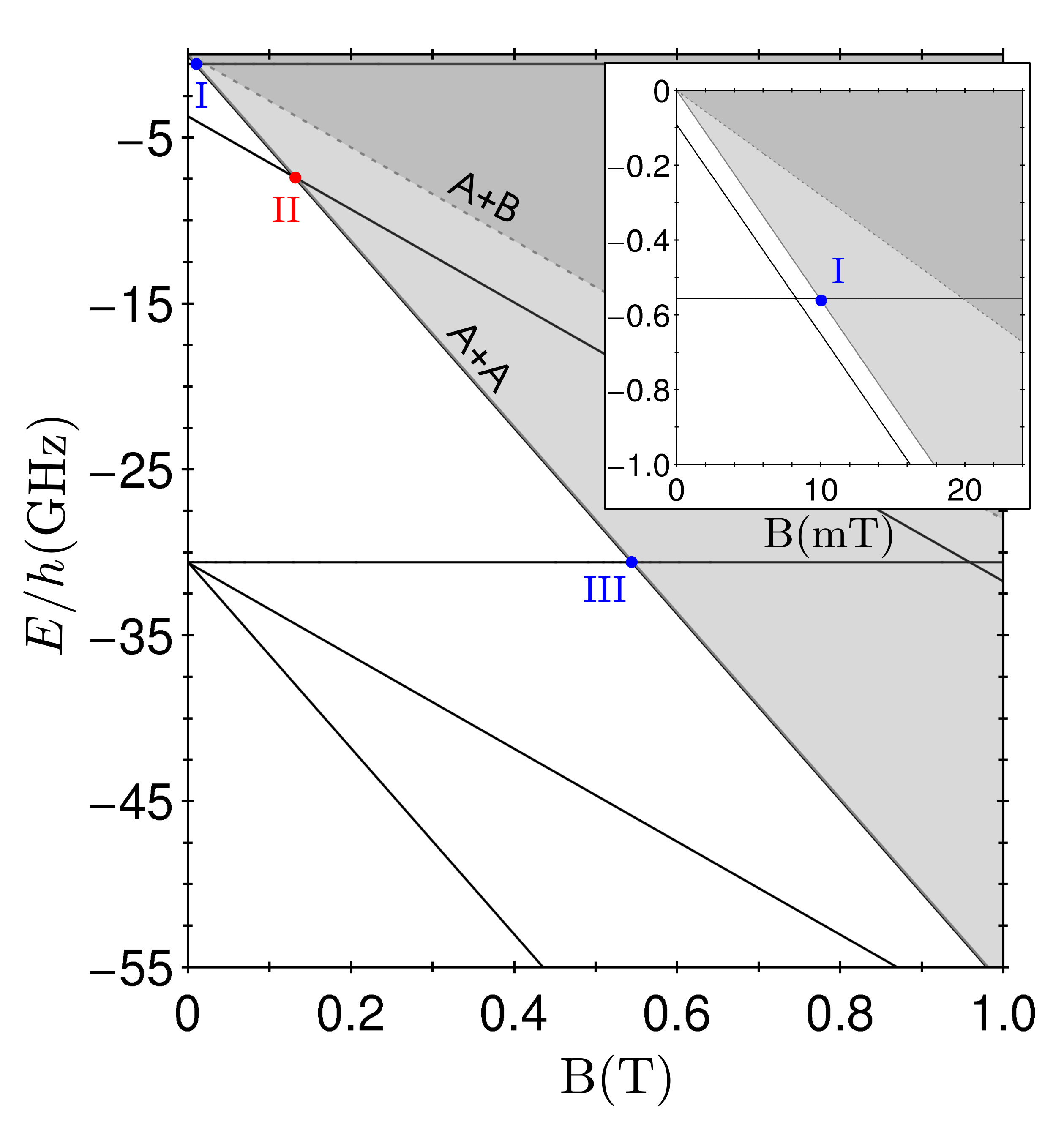}
\caption{(Color online) Energies of coupled bound states (black curve) are plotted versus the external magnetic field for $^4$He*--$^4$He*, neglecting threshold effects. The energy of the $\mt{A+A,A+B}$ states correspond to different channel threshold energies (gray curve). At the points $\mt{I,II,III}$ a coupled bound state intersects with the $\mt{A+A}$ threshold. The threshold of $\mt{A+A}$ is degenerate with a $d$-wave $\mt{A+A}$ state. For the points $\mt{I,III}$ Feshbach resonances are induced by $d$-wave bound states, with $S=0,2$ spin quantum numbers respectively. For $\mt{I}$ there will be a large spread in possible resonance field position $B_0$ because of the inaccuracy of the singlet potential. At $\mt{II}$ a $p$-wave bound state crosses the threshold but this state cannot induce a Feshbach resonance as it is not coupled the entrance channel.  
 }\label{fig:He44Vdd}
\end{figure}

A $d$-wave singlet bound state couples the entrance channel via $\mc{V}_{dd}$ and causes a narrow Feshbach resonance at $B_0=9.9$~mT whose (field) width equals $\Delta B=0.2$~$\mu$T which is denoted as $\mt{I}$ in Fig.~\ref{fig:He44Vdd}. If we make the singlet potential deeper, within the inaccuracy of the potential, we may find this resonance for fields up to $B_0=46$~mT with the same field width. By making the potential more shallow, in comparision with the nominal potential, the resonance will be found at lower magnetic fields and can even dissappear. Around $B=133.1$~mT a $p$-wave triplet bound state will cross the threshold of $\mt{A+A}$ (denoted by $\mt{II}$). This bound state can however not couple to the open channels and will therefore not cause a Feshbach resonance. For higher magnetic fields, a Feshbach resonance is found at $B_0=546.0$~mT and a field width of $\Delta B=1.1$~$\mu$T. This resonance is caused by a $d$-wave quintet bound state which is shown at $\mt{III}$ in Fig.~\ref{fig:He44Vdd}. Since the $S=2$ potential is the most accurate potential we have, we expect that the prediction of the $\mt{III}$ resonance will be the most accurate, as opposed to $\mt{I}$ where an $S=0$ bound state causes the resonance. For the $\mt{C+C}$ channel, Feshbach resonances are absent. We note however that within the uncertainty of the singlet potential it is possible for a $S=0,\ell=2$ bound state to be relatively close to the $S=2,\ell=0$ bound state at low magnetic fields. Since these two bound states can be coupled by the $\mc{V}_{dd}$ interaction, the $S=2,\ell=0$ bound state will become less stable. This can be of interest to recent studies~\cite{beams:06, moal:07} on the lifetime of the $S=2,\ell=0$ bound state of the $\mt{C+C}$ channel. This qualitative argument needs further investigation. 

For two fermionic helium atoms colliding, the entrance channels $\mt{a+a,d+d}$ can only be populated for $\ell=$odd. Therefore, for ultracold scattering experiments, the only relevant $s$-wave channel is $\mt{a+b}$. For this (anti-symmetrized) state the scattering is dominated by the quintet potential at high magnetic fields, see Fig.~\ref{fig:qfrac}. Without threshold effects, we expect that around $B\approx 970$~mT an $s$-wave quintet bound state will cross the scattering threshold. Unlike the Feshbach resonances discussed so-far, the energy of this coupled bound state is higher than the threshold energy for low magnetic fields. Including threshold effects we find that this bound state will not cause a Feshbach resonance. There is an $s$-wave triplet bound state which does create a wide ($\Delta B\approx 15.2$~mT) Feshbach resonance around $B_0\approx 1426.5$~mT. The magnetic dipole-dipole interaction induces five Feshbach resonances which are all caused by $d$-wave bound states in either the singlet or triplet configuration, hence these resonance positions will not be stated here.

Since the $S=1$ potential is almost resonant within the uncertainty variations, it may be able to capture a new bound state, see the inset of Fig.~\ref{fig:bindingenergy}. This makes it very challenging to reliably predict Feshbach resonances for $^3$He*-$^3$He* that involve the $S=1$ potential. 

\subsection*{Heteronuclear gas}

For the heteronuclear gas the only selected channel which allows for Feshbach resonances induced by the central interaction is $\mt{b+A}$. The $\mt{a+A}$ and $\mt{d+C}$ mixtures are both fully-stretched states, and therefore can only have Feshbach resonances induced by $\mc{V}_{dd}$. If we apply the ABM without taking into account threshold effects, we find two resonances: one at low field $B_0=1.3$~mT and one $B_0=347.8$~mT. However, since the least bound quintet state ($\epsilon_2^{34}$) is almost resonant, threshold effects will dominate and broad Feshbach resonances are expected. 

\begin{figure}[!h]
\centering
\includegraphics[width=\columnwidth,keepaspectratio=true]{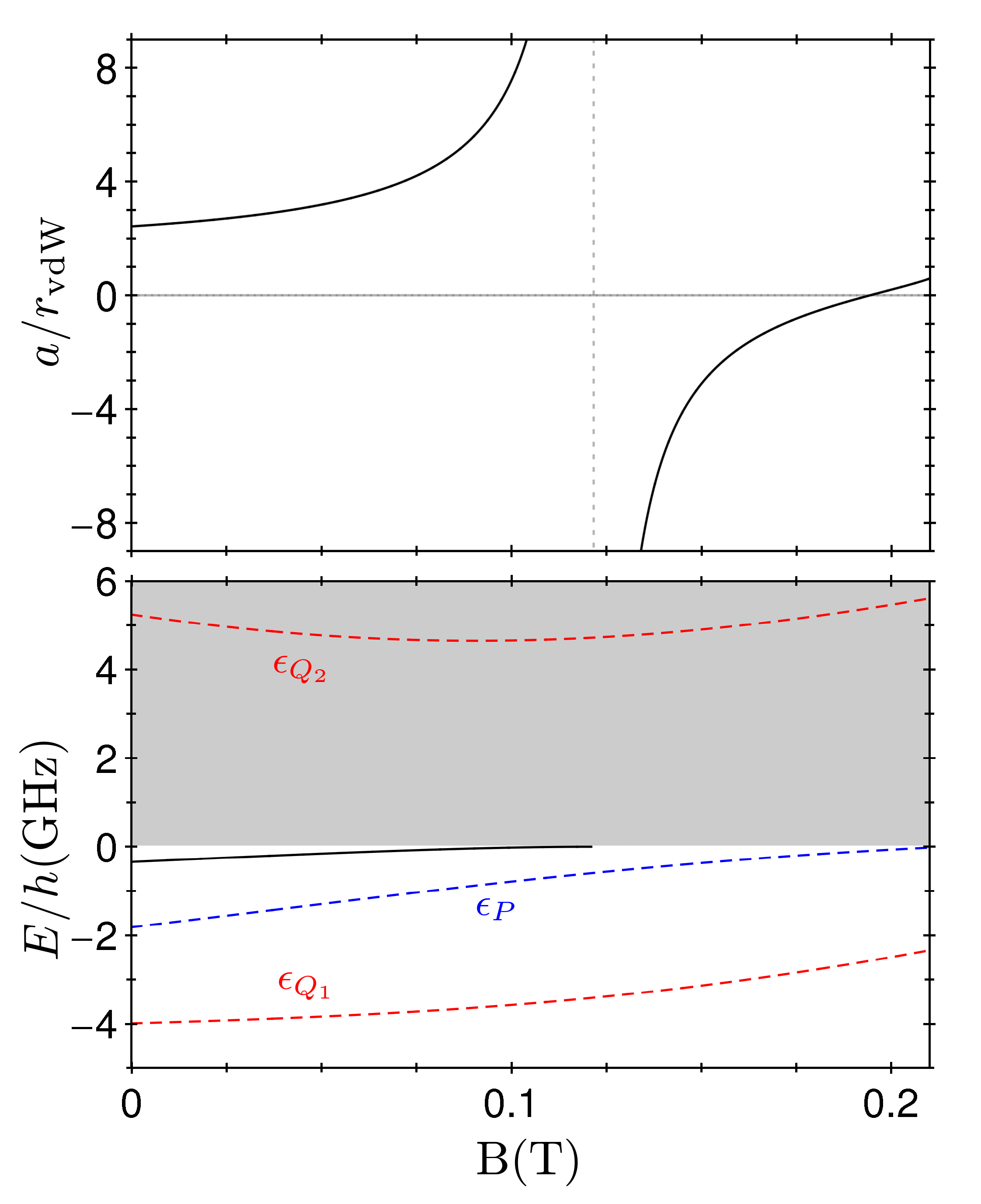}
\caption{(Color online) In the upper panel the scattering length is shown for the $\mt{b+A}$ state as a function of magnetic field. The scattering length is scaled to the van der Waals range $r_\mt{vdW}= 34\,\mathrm{a_0} $. At a magnetic field of $B_0=121.4$~mT the Feshbach resonance will occur, which has a field width of $\Delta B=72.9$~mT. At a magnetic field of $B_0^{P}\approx214.7$~mT, just outside the plot, the $\mc{P}$-space bound state will become resonant. In the lower panel the energies of the uncoupled $\mc P,Q$-space bound states are shown (dashed lines blue $\epsilon_P$ and red curves $\epsilon_{Q_1},\epsilon_{Q_2}$ respectively) with respect to the open channel threshold energy. The energy of the dressed bound state, which causes the Feshbach resonance, is found by solving the pole equation of the $S$-matrix (black solid curve).}\label{fig:He34FR}
\end{figure}

To include threshold effects we apply the theory as described in Sect.~\ref{sect:ABM}. We determine the uncoupled $\mc{P,Q}$ space bound states. Remarkably we do not find that either of the bare closed channel bound states (with energies $\epsilon_{Q_1}$ and $\epsilon_{Q_2}$) cross the scattering threshold at low magnetic field (see Fig.~\ref{fig:He34FR} ), as one would usually expect. Counter intuitively it is the energy of the bare open channel bound state $\epsilon_P$ that becomes degenerate with the scattering threshold which, if we could physically uncouple $\mc{P}$- and $\mc{Q}$-space ($\mc{H}_{PQ}\to0$), would result in a potential resonance. This will have severe consequences for the observed resonance structure. Where usually a $\mc{Q}$-space bound state pushes the dressed bound state through threshold, here it is the $\mc{P}$-space bound state. 

The single resonance approximation will now fail since the  $|\phi_{Q_1}\ket$  and $|\phi_{Q_2}\ket$ states are energetically almost equidistant to the threshold. To describe the dressed bound state of the coupled system we need to study the peculiar interplay of three bare bound states: $|\phi_{P}\ket,|\phi_{Q_1}\ket,|\phi_{Q_2}\ket$. The interplay of these bound states is illustrated nicely by the pole equation of the total scattering matrix $S=S_PS_Q$: 
\begin{equation}\label{eq:poleEq}
(\kappa_P+ik)(E-E_1)(E-E_2)=0,
\end{equation}
where we have used Eq.~(\ref{eqn:SQ}) for $S_Q$. The closed channel bound states $|\phi_{Q_1}\ket,|\phi_{Q_2}\ket$ not only interact with the $|\phi_P\ket$ state but also with each other via $\mathcal{P}$-space. The energy of the dressed bound state which results from this interplay is shown as a solid line in Fig.~\ref{fig:He34FR}.  Here only the physical solution of Eq.~(\ref{eq:poleEq}) which causes the resonance is shown.

From these threshold effects, we find a Feshbach resonance position $B_0=121.4^{+52.7}_{-45.9}$~mT and width $\Delta B=72.9^{+18.3}_{-19.3}$~mT. The variation in the position and width of the resonance stated here is due to the uncertainty in the triplet bound state, since a significant fraction of the $|\phi_{Q_1}\ket$ state is in a triplet state. The strong coupling between open and closed channel states (large $\mc{H}_{PQ}$) in combination with the $\mathcal{P}$-space bound state being almost resonant, results in this very broad Feshbach resonance.

The open channel bound state crosses the threshold at a magnetic field of $B_0^{P}=214.7$~mT. The description of the binding energies and the scattering length for $B>B_0^{P}$ will become inaccurate. 

\begin{figure}[!h]
\centering
\includegraphics[width=\columnwidth,keepaspectratio=true]{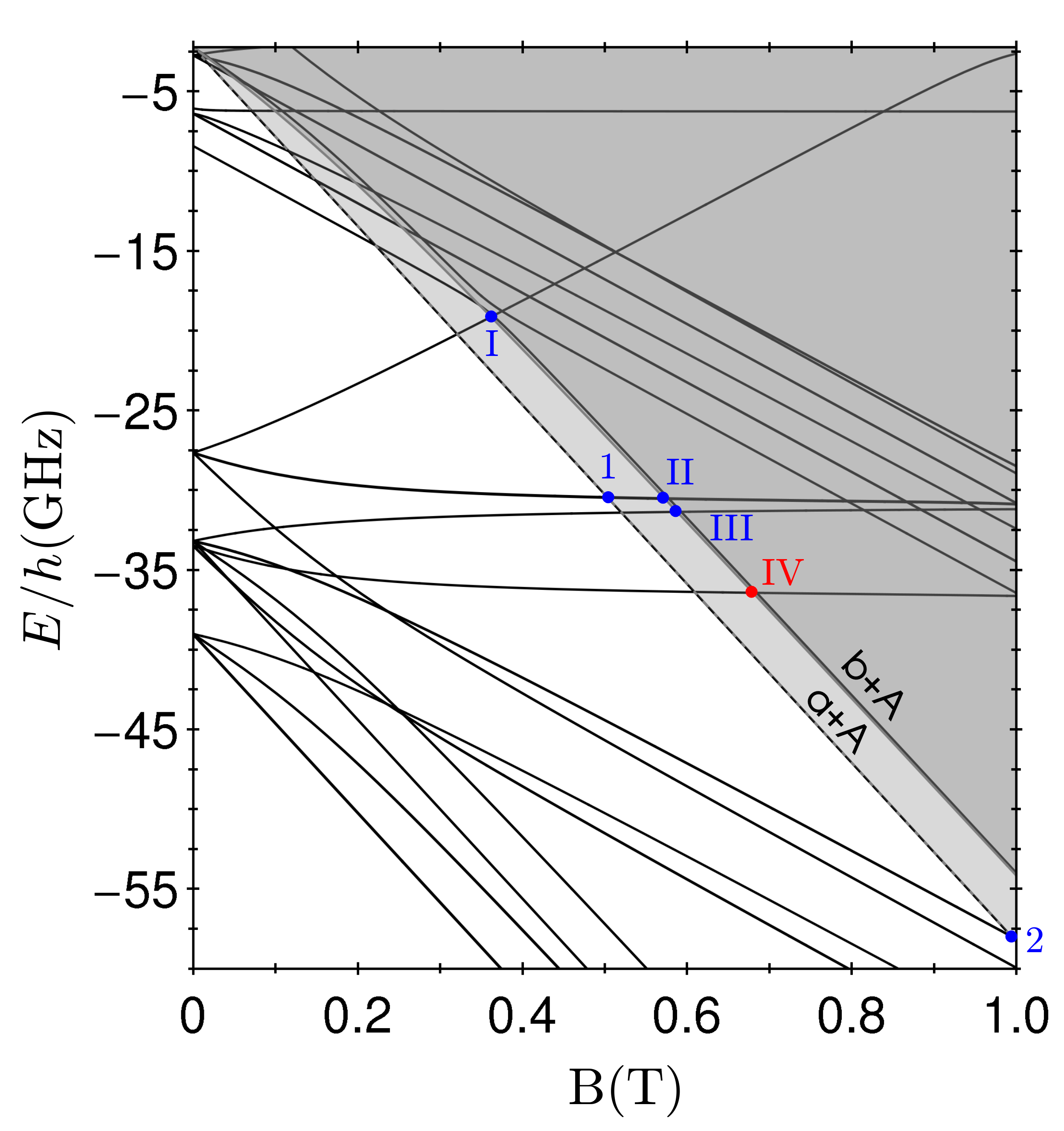}
\caption{(Color online) Coupled bound states diagram for the $^3$He*--$^4$He* mixture in the $\mt{a+A,b+A}$ states. Only the positions of Feshbach resonances induced by $S=2$ bound states are labeled and the precise values are given in the text. 
For the spin stretched entrance channel $\mt{a+A}$, we find two resonances induced by ($d$-wave) quintet bound states at (1,2). For the $\mt{b+A}$ entrance channel quintet ($d$-wave) bound states cause Feshbach resonances at $\mt{I,II,III}$. For $\mt{IV}$ a $p$-wave bound state crosses the threshold. This bound state can only be coupled to other states of $\ell$=odd, and will thus not couple to the threshold $s$-wave state. All other crossings correspond to coupled bound states of $S<2$.  Note the large number of possible resonances at low magnetic fields.}\label{fig:He34Vdd}
\end{figure}

The magnetic dipole-dipole interaction for the $\mt{b+A}$ state can induce multiple Feshbach resonances. Since the $\mt{a+A}$ channel is energetically open when we include $\mc{V}_{dd}$ we expect the Feshbach resonances to be modified by these open channels. Therefore we will only predict the positions (thus without threshold effects) of the three resonances $B_0=\{361.8,572.9,587.9\}$~mT induced by quintet $d$-wave bound states, $\mt{I,II,III}$ respectively in Fig.~\ref{fig:He34Vdd}. At $\mt{IV}$ a $p$-wave quintet bound state will cross the threshold. This bound state cannot couple to $s$-wave scattering states in $\mt{b+A}$. 

In the fully stretched $\mt{a+A,d+C}$ states, only $\mc{V}_{dd}$ induced Feshbach resonances can occur. For $\mt{a+A}$ we find five Feshbach resonances. Two of those are caused by quintet ($d$-wave) bound states which yield $B_0=503.0$~mT, $\Delta B=5.5$~$\mu$T and $B_0=994.1$~mT, $\Delta B=0.1$~$\mu$T labeled as $1,2$ respectively in Fig.~\ref{fig:He34Vdd}. The $\mt{d+C}$ entrance channel state is expected to suffer more from inelastic losses due to $\mc{V}_{dd}$ as there are multiple (non-degenerate) open channels. We only find one Feshbach resonance for this channel which is caused by a triplet $d$-wave bound state. 

\begin{table}
\caption{\label{tab:FRsummary} A summary of the predicted Feshbach resonances for He*. The accuracy of the predicted resonance field position ($B_0$) is mainly determined by the dominant $S$ value of the responsible coupled bound state. All error bars are based on the deviations in the used parameters, not on the errors made by applying the ABM. $^1$) For $\mc{A}=44$ the resonance at $B_0=9.9$~mT can, within its error bars, shift up to $B_0=46$~mT and below $B_0=0$~mT, i.e.~disappear.
$^2$) For $\mc{A}=33$ the nominal result is stated, although, due to the resonant nature of the $S=1$ potential, the resonance structure can change dramatically. This will lead to a large deviation ($\approx 300$~mT) in $B_0$. }
\begin{tabular}{cccccccc}
\hline
\hline
$\mc{A}$ & mixture & $B_0$~(mT) & $\Delta B$~(mT) & $\ell$ & $S$ & coupling\\
\hline \\ [-2ex]
44 & A+A & $9.9^1$ & 0.2$\times 10^{-3}$ & 2 & 0 & $\mathcal{V}_{dd}$\\ [.5ex]
44 & A+A & 546.0(1) & 1$\times 10^{-3}$ & 2 & 2 & $\mathcal{V}_{dd}$\\ [.5ex]
\hline \\ [-2ex]
33 & a+b & $1426.5^2$ & 13.2 & 0 & 1 & $\mathcal{V}$\\ [.5ex]
\hline \\ [-2ex]
34 & b+A & $121.4^{+52.7}_{-45.9}$ & $72.9^{+18.3}_{-19.3}$ & 0 & 2& $\mathcal{V}$\\ [.5ex]
34 & b+A & 361.8(6) & - & 2 & 2& $\mathcal{V}_{dd}$\\ [.5ex]
34 & b+A & 572.9(1) & - & 2 & 2& $\mathcal{V}_{dd}$\\ [.5ex]
34 & b+A & 587.9(1) & - & 2 & 2& $\mathcal{V}_{dd}$\\ [.5ex]
34 & a+A & 503.0(1) & 5.5$\times 10^{-3}$ & 2 & 2& $\mathcal{V}_{dd}$\\ [.5ex]
34 & a+A & 994.1(1) & 0.1$\times 10^{-3}$ & 2 & 2& $\mathcal{V}_{dd}$\\ [.5ex]
\hline
\hline
\end{tabular}
\end{table}

\section{Concluding remarks}

We presented here the first study of Feshbach resonances for metastable helium atoms, by using the Asymptotic Bound-state Model. By analyzing the various inelastic decay processes we have selected a few two-body spin states suitable for observing Feshbach resonances. Reliably predicting these resonances is in some cases hindered by fact that the $S=0,1$ potentials are known with far less accuracy as compared to the $S=2$ molecular potential. Therefore, we have limited our discussion to coupled bound states which have a dominant quintet character and cause Feshbach resonances for magnetic fields up to $1$~T. To study these resonances we have utilized and expanded the ABM: magnetic dipole-dipole interactions as well as overlapping resonances can now be described with this model. Although we found several Feshbach resonances, we did not find wide resonances for the homonuclear (bosonic and fermionic) gas for the selected spin states. The heteronuclear system however, reveals a very wide resonance $\Delta B=72.9~\mathrm{mT}$ at relatively low magnetic field $B_0=121.4~\mathrm{mT}$ making it of potential interest for further theoretical and experimental investigation. 

Measurements of Feshbach resonances will aid enormously in constructing more accurate $S=0,1$ molecular potentials. Vice-versa, with more accurate $S=0,1$ interaction potentials we will be able to give reliable predictions of the abundant number of resonances induced by coupled bound states in a dominantly singlet or triplet state. It is our current understanding that these potentials can induce resonances at lower magnetic fields as compared to $S=2$ dominated bound states which makes them of great potential interest. Accurate interaction potentials would pave the way for full, numerically exact, coupled channels calculations which will yield more accurate predictions. Based on qualitative arguments, we also point out the interesting possibility of Feshbach resonance induced stabilization of PI losses, where the resonance can effectively reduce the inelastic loss rate. This effect may for instance be used to stabilize $S=0,1$ (coupled) bound states. 

TGT acknowledges support of the research program on Quantum Gases of the Stichting voor Fundamenteel Onderzoek der Materie (FOM), which is financially supported by the Nederlandse Organisatie voor Wetenschappelijk Onderzoek (NWO). WV acknowledges support from the FOM.

\bibliography{bibHeFR}

\begin{thebibliography}{46}
\expandafter\ifx\csname natexlab\endcsname\relax\def\natexlab#1{#1}\fi
\expandafter\ifx\csname bibnamefont\endcsname\relax
  \def\bibnamefont#1{#1}\fi
\expandafter\ifx\csname bibfnamefont\endcsname\relax
  \def\bibfnamefont#1{#1}\fi
\expandafter\ifx\csname citenamefont\endcsname\relax
  \def\citenamefont#1{#1}\fi
\expandafter\ifx\csname url\endcsname\relax
  \def\url#1{\texttt{#1}}\fi
\expandafter\ifx\csname urlprefix\endcsname\relax\def\urlprefix{URL }\fi
\providecommand{\bibinfo}[2]{#2}
\providecommand{\eprint}[2][]{\url{#2}}

\bibitem[{\citenamefont{M\"{u}ller et~al.}(1991)\citenamefont{M\"{u}ller, Merz,
  Ruf, Hotop, Meyer, and Movre}}]{muller:91}
\bibinfo{author}{\bibfnamefont{M.~W.} \bibnamefont{M\"{u}ller}},
  \bibinfo{author}{\bibfnamefont{A.}~\bibnamefont{Merz}},
  \bibinfo{author}{\bibfnamefont{M.~W.} \bibnamefont{Ruf}},
  \bibinfo{author}{\bibfnamefont{H.}~\bibnamefont{Hotop}},
  \bibinfo{author}{\bibfnamefont{W.}~\bibnamefont{Meyer}}, \bibnamefont{and}
  \bibinfo{author}{\bibfnamefont{M.}~\bibnamefont{Movre}},
  \bibinfo{journal}{Zeitschrift f\"{u}r Physik D}
  \textbf{\bibinfo{volume}{21}}, \bibinfo{pages}{89} (\bibinfo{year}{1991}).

\bibitem[{\citenamefont{St\"{a}rck and Meyer}(1994)}]{starck:94}
\bibinfo{author}{\bibfnamefont{J.}~\bibnamefont{St\"{a}rck}} \bibnamefont{and}
  \bibinfo{author}{\bibfnamefont{W.}~\bibnamefont{Meyer}},
  \bibinfo{journal}{Chem. Phys. lett.} \textbf{\bibinfo{volume}{225}},
  \bibinfo{pages}{229} (\bibinfo{year}{1994}).

\bibitem[{\citenamefont{Dickinson et~al.}(2004)\citenamefont{Dickinson, Gadea,
  and Leininger}}]{dickinson:04}
\bibinfo{author}{\bibfnamefont{A.~S.} \bibnamefont{Dickinson}},
  \bibinfo{author}{\bibfnamefont{F.~X.} \bibnamefont{Gadea}}, \bibnamefont{and}
  \bibinfo{author}{\bibfnamefont{T.}~\bibnamefont{Leininger}},
  \bibinfo{journal}{J. Phys. B-At. Mol. Opt. Phys.}
  \textbf{\bibinfo{volume}{37}}, \bibinfo{pages}{587} (\bibinfo{year}{2004}).

\bibitem[{\citenamefont{Przybytek and Jeziorski}(2005)}]{przybytek:05}
\bibinfo{author}{\bibfnamefont{M.}~\bibnamefont{Przybytek}} \bibnamefont{and}
  \bibinfo{author}{\bibfnamefont{B.}~\bibnamefont{Jeziorski}},
  \bibinfo{journal}{J. Chem. Phys.} \textbf{\bibinfo{volume}{123}},
  \bibinfo{pages}{134315} (\bibinfo{year}{2005}).

\bibitem[{\citenamefont{Robert et~al.}(2001)\citenamefont{Robert, Sirjean,
  Browaeys, Poupard, Nowak, Boiron, Westbrook, and Aspect}}]{robert:01}
\bibinfo{author}{\bibfnamefont{A.}~\bibnamefont{Robert}},
  \bibinfo{author}{\bibfnamefont{O.}~\bibnamefont{Sirjean}},
  \bibinfo{author}{\bibfnamefont{A.}~\bibnamefont{Browaeys}},
  \bibinfo{author}{\bibfnamefont{J.}~\bibnamefont{Poupard}},
  \bibinfo{author}{\bibfnamefont{S.}~\bibnamefont{Nowak}},
  \bibinfo{author}{\bibfnamefont{D.}~\bibnamefont{Boiron}},
  \bibinfo{author}{\bibfnamefont{C.~I.} \bibnamefont{Westbrook}},
  \bibnamefont{and} \bibinfo{author}{\bibfnamefont{A.}~\bibnamefont{Aspect}},
  \bibinfo{journal}{Science} \textbf{\bibinfo{volume}{292}},
  \bibinfo{pages}{461} (\bibinfo{year}{2001}).

\bibitem[{\citenamefont{Pereira Dos~Santos et~al.}(2001)\citenamefont{Pereira
  Dos~Santos, L\'eonard, Wang, Barrelet, Perales, Rasel, Unnikrishnan, Leduc,
  and Cohen-Tannoudji}}]{santos:01}
\bibinfo{author}{\bibfnamefont{F.}~\bibnamefont{Pereira Dos~Santos}},
  \bibinfo{author}{\bibfnamefont{J.}~\bibnamefont{L\'eonard}},
  \bibinfo{author}{\bibfnamefont{J.}~\bibnamefont{Wang}},
  \bibinfo{author}{\bibfnamefont{C.~J.} \bibnamefont{Barrelet}},
  \bibinfo{author}{\bibfnamefont{F.}~\bibnamefont{Perales}},
  \bibinfo{author}{\bibfnamefont{E.}~\bibnamefont{Rasel}},
  \bibinfo{author}{\bibfnamefont{C.~S.} \bibnamefont{Unnikrishnan}},
  \bibinfo{author}{\bibfnamefont{M.}~\bibnamefont{Leduc}}, \bibnamefont{and}
  \bibinfo{author}{\bibfnamefont{C.}~\bibnamefont{Cohen-Tannoudji}},
  \bibinfo{journal}{Phys. Rev. Lett.} \textbf{\bibinfo{volume}{86}},
  \bibinfo{pages}{3459} (\bibinfo{year}{2001}).

\bibitem[{\citenamefont{Tychkov et~al.}(2006)\citenamefont{Tychkov, Jeltes,
  McNamara, Tol, Herschbach, Hogervorst, and Vassen}}]{tychkov:06}
\bibinfo{author}{\bibfnamefont{A.~S.} \bibnamefont{Tychkov}},
  \bibinfo{author}{\bibfnamefont{T.}~\bibnamefont{Jeltes}},
  \bibinfo{author}{\bibfnamefont{J.~M.} \bibnamefont{McNamara}},
  \bibinfo{author}{\bibfnamefont{P.~J.~J.} \bibnamefont{Tol}},
  \bibinfo{author}{\bibfnamefont{N.}~\bibnamefont{Herschbach}},
  \bibinfo{author}{\bibfnamefont{W.}~\bibnamefont{Hogervorst}},
  \bibnamefont{and} \bibinfo{author}{\bibfnamefont{W.}~\bibnamefont{Vassen}},
  \bibinfo{journal}{Phys. Rev. A} \textbf{\bibinfo{volume}{73}},
  \bibinfo{pages}{031603} (\bibinfo{year}{2006}).

\bibitem[{\citenamefont{Dall and Truscott}(2007)}]{dall:07}
\bibinfo{author}{\bibfnamefont{R.~G.} \bibnamefont{Dall}} \bibnamefont{and}
  \bibinfo{author}{\bibfnamefont{A.~G.} \bibnamefont{Truscott}},
  \bibinfo{journal}{Opt. Commun.} \textbf{\bibinfo{volume}{270}},
  \bibinfo{pages}{255} (\bibinfo{year}{2007}).

\bibitem[{\citenamefont{Doret et~al.}(2009)\citenamefont{Doret, Connolly,
  Ketterle, and Doyle}}]{ketterle:09}
\bibinfo{author}{\bibfnamefont{S.~C.} \bibnamefont{Doret}},
  \bibinfo{author}{\bibfnamefont{C.~B.} \bibnamefont{Connolly}},
  \bibinfo{author}{\bibfnamefont{W.}~\bibnamefont{Ketterle}}, \bibnamefont{and}
  \bibinfo{author}{\bibfnamefont{J.~M.} \bibnamefont{Doyle}},
  \bibinfo{journal}{Phys. Rev. Lett.} \textbf{\bibinfo{volume}{103}},
  \bibinfo{pages}{103005} (\bibinfo{year}{2009}).

\bibitem[{\citenamefont{McNamara et~al.}(2006)\citenamefont{McNamara, Jeltes,
  Tychkov, Hogervorst, and Vassen}}]{mcnamara:06}
\bibinfo{author}{\bibfnamefont{J.~M.} \bibnamefont{McNamara}},
  \bibinfo{author}{\bibfnamefont{T.}~\bibnamefont{Jeltes}},
  \bibinfo{author}{\bibfnamefont{A.~S.} \bibnamefont{Tychkov}},
  \bibinfo{author}{\bibfnamefont{W.}~\bibnamefont{Hogervorst}},
  \bibnamefont{and} \bibinfo{author}{\bibfnamefont{W.}~\bibnamefont{Vassen}},
  \bibinfo{journal}{Phys. Rev. Lett.} \textbf{\bibinfo{volume}{97}},
  \bibinfo{pages}{080404} (\bibinfo{year}{2006}).

\bibitem[{\citenamefont{Chin et~al.}(2010)\citenamefont{Chin, Grimm, Julienne,
  and Tiesinga}}]{chin:10}
\bibinfo{author}{\bibfnamefont{C.}~\bibnamefont{Chin}},
  \bibinfo{author}{\bibfnamefont{R.}~\bibnamefont{Grimm}},
  \bibinfo{author}{\bibfnamefont{P.}~\bibnamefont{Julienne}}, \bibnamefont{and}
  \bibinfo{author}{\bibfnamefont{E.}~\bibnamefont{Tiesinga}},
  \bibinfo{journal}{Rev. Mod. Phys.} \textbf{\bibinfo{volume}{82}},
  \bibinfo{pages}{1225} (\bibinfo{year}{2010}).

\bibitem[{\citenamefont{M\o{}lmer}(1998)}]{molmer:98}
\bibinfo{author}{\bibfnamefont{K.}~\bibnamefont{M\o{}lmer}},
  \bibinfo{journal}{Phys. Rev. Lett.} \textbf{\bibinfo{volume}{80}},
  \bibinfo{pages}{1804} (\bibinfo{year}{1998}).

\bibitem[{\citenamefont{Efremov and Viverit}(2002)}]{efremov:02}
\bibinfo{author}{\bibfnamefont{D.~V.} \bibnamefont{Efremov}} \bibnamefont{and}
  \bibinfo{author}{\bibfnamefont{L.}~\bibnamefont{Viverit}},
  \bibinfo{journal}{Phys. Rev. B} \textbf{\bibinfo{volume}{65}},
  \bibinfo{pages}{134519} (\bibinfo{year}{2002}).

\bibitem[{\citenamefont{Schellekens et~al.}(2005)\citenamefont{Schellekens,
  Hoppeler, Perrin, Viana~Gomes, Boiron, Aspect, and
  Westbrook}}]{schellekens:05}
\bibinfo{author}{\bibfnamefont{M.}~\bibnamefont{Schellekens}},
  \bibinfo{author}{\bibfnamefont{R.}~\bibnamefont{Hoppeler}},
  \bibinfo{author}{\bibfnamefont{A.}~\bibnamefont{Perrin}},
  \bibinfo{author}{\bibfnamefont{J.}~\bibnamefont{Viana~Gomes}},
  \bibinfo{author}{\bibfnamefont{D.}~\bibnamefont{Boiron}},
  \bibinfo{author}{\bibfnamefont{A.}~\bibnamefont{Aspect}}, \bibnamefont{and}
  \bibinfo{author}{\bibfnamefont{C.}~\bibnamefont{Westbrook}},
  \bibinfo{journal}{Science} \textbf{\bibinfo{volume}{310}},
  \bibinfo{pages}{648} (\bibinfo{year}{2005}).

\bibitem[{\citenamefont{Jeltes et~al.}(2007)\citenamefont{Jeltes, McNamara,
  Hogervorst, Vassen, Krachmalnicoff, Schellekens, Perrin, Chang, Boiron,
  Aspect et~al.}}]{jeltes:07}
\bibinfo{author}{\bibfnamefont{T.}~\bibnamefont{Jeltes}},
  \bibinfo{author}{\bibfnamefont{J.~M.} \bibnamefont{McNamara}},
  \bibinfo{author}{\bibfnamefont{W.}~\bibnamefont{Hogervorst}},
  \bibinfo{author}{\bibfnamefont{W.}~\bibnamefont{Vassen}},
  \bibinfo{author}{\bibfnamefont{V.}~\bibnamefont{Krachmalnicoff}},
  \bibinfo{author}{\bibfnamefont{M.}~\bibnamefont{Schellekens}},
  \bibinfo{author}{\bibfnamefont{A.}~\bibnamefont{Perrin}},
  \bibinfo{author}{\bibfnamefont{H.}~\bibnamefont{Chang}},
  \bibinfo{author}{\bibfnamefont{D.}~\bibnamefont{Boiron}},
  \bibinfo{author}{\bibfnamefont{A.}~\bibnamefont{Aspect}},
  \bibnamefont{et~al.}, \bibinfo{journal}{Nature}
  \textbf{\bibinfo{volume}{445}}, \bibinfo{pages}{402} (\bibinfo{year}{2007}).

\bibitem[{\citenamefont{Perrin et~al.}(2007)\citenamefont{Perrin, Chang,
  Krachmalnicoff, Schellekens, Boiron, Aspect, and Westbrook}}]{perrin:07}
\bibinfo{author}{\bibfnamefont{A.}~\bibnamefont{Perrin}},
  \bibinfo{author}{\bibfnamefont{H.}~\bibnamefont{Chang}},
  \bibinfo{author}{\bibfnamefont{V.}~\bibnamefont{Krachmalnicoff}},
  \bibinfo{author}{\bibfnamefont{M.}~\bibnamefont{Schellekens}},
  \bibinfo{author}{\bibfnamefont{D.}~\bibnamefont{Boiron}},
  \bibinfo{author}{\bibfnamefont{A.}~\bibnamefont{Aspect}}, \bibnamefont{and}
  \bibinfo{author}{\bibfnamefont{C.~I.} \bibnamefont{Westbrook}},
  \bibinfo{journal}{Phys. Rev. Lett.} \textbf{\bibinfo{volume}{99}},
  \bibinfo{eid}{150405} (\bibinfo{year}{2007}).

\bibitem[{\citenamefont{Przybytek}(2008)}]{przybytek:08}
\bibinfo{author}{\bibfnamefont{M.}~\bibnamefont{Przybytek}}, Ph.D. thesis,
  \bibinfo{school}{University of Warsaw, Poland} (\bibinfo{year}{2008}).

\bibitem[{\citenamefont{Moal et~al.}(2006)\citenamefont{Moal, Portier, Kim,
  Dugu\'e, Rapol, Leduc, and Cohen-Tannoudji}}]{moal:06}
\bibinfo{author}{\bibfnamefont{S.}~\bibnamefont{Moal}},
  \bibinfo{author}{\bibfnamefont{M.}~\bibnamefont{Portier}},
  \bibinfo{author}{\bibfnamefont{J.}~\bibnamefont{Kim}},
  \bibinfo{author}{\bibfnamefont{J.}~\bibnamefont{Dugu\'e}},
  \bibinfo{author}{\bibfnamefont{U.~D.} \bibnamefont{Rapol}},
  \bibinfo{author}{\bibfnamefont{M.}~\bibnamefont{Leduc}}, \bibnamefont{and}
  \bibinfo{author}{\bibfnamefont{C.}~\bibnamefont{Cohen-Tannoudji}},
  \bibinfo{journal}{Phys. Rev. Lett.} \textbf{\bibinfo{volume}{96}},
  \bibinfo{pages}{023203} (\bibinfo{year}{2006}).

\bibitem[{\citenamefont{Partridge et~al.}(2010)\citenamefont{Partridge,
  Jaskula, Bonneau, Boiron, and Westbrook}}]{partridge:10}
\bibinfo{author}{\bibfnamefont{G.~B.} \bibnamefont{Partridge}},
  \bibinfo{author}{\bibfnamefont{J.-C.} \bibnamefont{Jaskula}},
  \bibinfo{author}{\bibfnamefont{M.}~\bibnamefont{Bonneau}},
  \bibinfo{author}{\bibfnamefont{D.}~\bibnamefont{Boiron}}, \bibnamefont{and}
  \bibinfo{author}{\bibfnamefont{C.~I.} \bibnamefont{Westbrook}},
  \bibinfo{journal}{Phys. Rev. A} \textbf{\bibinfo{volume}{81}},
  \bibinfo{pages}{053631} (\bibinfo{year}{2010}).

\bibitem[{\citenamefont{Tiecke et~al.}(2010{\natexlab{a}})\citenamefont{Tiecke,
  Goosen, Walraven, and Kokkelmans}}]{tiecke:10b}
\bibinfo{author}{\bibfnamefont{T.~G.} \bibnamefont{Tiecke}},
  \bibinfo{author}{\bibfnamefont{M.~R.} \bibnamefont{Goosen}},
  \bibinfo{author}{\bibfnamefont{J.~T.~M.} \bibnamefont{Walraven}},
  \bibnamefont{and} \bibinfo{author}{\bibfnamefont{S.~J.~J.~M.~F.}
  \bibnamefont{Kokkelmans}}, \bibinfo{journal}{arXiv:1007.0886}
  (\bibinfo{year}{2010}{\natexlab{a}}).

\bibitem[{\citenamefont{Verhaar et~al.}(2009)\citenamefont{Verhaar, van Kempen,
  and Kokkelmans}}]{verhaar:09}
\bibinfo{author}{\bibfnamefont{B.~J.} \bibnamefont{Verhaar}},
  \bibinfo{author}{\bibfnamefont{E.~G.~M.} \bibnamefont{van Kempen}},
  \bibnamefont{and} \bibinfo{author}{\bibfnamefont{S.~J. J. M.~F.}
  \bibnamefont{Kokkelmans}}, \bibinfo{journal}{Physical Review A (Atomic,
  Molecular, and Optical Physics)} \textbf{\bibinfo{volume}{79}},
  \bibinfo{eid}{032711} (\bibinfo{year}{2009}).

\bibitem[{\citenamefont{Wille et~al.}(2008)\citenamefont{Wille, Spiegelhalder,
  Kerner, Naik, Trenkwalder, Hendl, Schreck, Grimm, Tiecke, Walraven
  et~al.}}]{wille:08}
\bibinfo{author}{\bibfnamefont{E.}~\bibnamefont{Wille}},
  \bibinfo{author}{\bibfnamefont{F.~M.} \bibnamefont{Spiegelhalder}},
  \bibinfo{author}{\bibfnamefont{G.}~\bibnamefont{Kerner}},
  \bibinfo{author}{\bibfnamefont{D.}~\bibnamefont{Naik}},
  \bibinfo{author}{\bibfnamefont{A.}~\bibnamefont{Trenkwalder}},
  \bibinfo{author}{\bibfnamefont{G.}~\bibnamefont{Hendl}},
  \bibinfo{author}{\bibfnamefont{F.}~\bibnamefont{Schreck}},
  \bibinfo{author}{\bibfnamefont{R.}~\bibnamefont{Grimm}},
  \bibinfo{author}{\bibfnamefont{T.~G.} \bibnamefont{Tiecke}},
  \bibinfo{author}{\bibfnamefont{J.~T.~M.} \bibnamefont{Walraven}},
  \bibnamefont{et~al.}, \bibinfo{journal}{Phys. Rev. Lett.}
  \textbf{\bibinfo{volume}{100}}, \bibinfo{pages}{053201}
  (\bibinfo{year}{2008}).

\bibitem[{\citenamefont{Tiecke et~al.}(2010{\natexlab{b}})\citenamefont{Tiecke,
  Goosen, Ludewig, Gensemer, Kraft, Kokkelmans, and Walraven}}]{tiecke:10a}
\bibinfo{author}{\bibfnamefont{T.~G.} \bibnamefont{Tiecke}},
  \bibinfo{author}{\bibfnamefont{M.~R.} \bibnamefont{Goosen}},
  \bibinfo{author}{\bibfnamefont{A.}~\bibnamefont{Ludewig}},
  \bibinfo{author}{\bibfnamefont{S.~D.} \bibnamefont{Gensemer}},
  \bibinfo{author}{\bibfnamefont{S.}~\bibnamefont{Kraft}},
  \bibinfo{author}{\bibfnamefont{S.~J. J. M.~F.} \bibnamefont{Kokkelmans}},
  \bibnamefont{and} \bibinfo{author}{\bibfnamefont{J.~T.~M.}
  \bibnamefont{Walraven}}, \bibinfo{journal}{Phys. Rev. Lett.}
  \textbf{\bibinfo{volume}{104}}, \bibinfo{pages}{053202}
  (\bibinfo{year}{2010}{\natexlab{b}}).

\bibitem[{\citenamefont{Varshalovich et~al.}(1988)\citenamefont{Varshalovich,
  Moskalev, and Khersonskii}}]{AngularBook:88}
\bibinfo{author}{\bibfnamefont{D.~A.} \bibnamefont{Varshalovich}},
  \bibinfo{author}{\bibfnamefont{A.~N.} \bibnamefont{Moskalev}},
  \bibnamefont{and} \bibinfo{author}{\bibfnamefont{V.~K.}
  \bibnamefont{Khersonskii}}, \emph{\bibinfo{title}{Quantum theory of angular
  momentum}} (\bibinfo{publisher}{Singapore : World Scientific},
  \bibinfo{year}{1988}).

\bibitem[{\citenamefont{Feshbach}(1958)}]{feshbach:58}
\bibinfo{author}{\bibfnamefont{H.}~\bibnamefont{Feshbach}},
  \bibinfo{journal}{Ann. Phys.} \textbf{\bibinfo{volume}{5}},
  \bibinfo{pages}{357} (\bibinfo{year}{1958}).

\bibitem[{\citenamefont{Feshbach}(1962)}]{feshbach:62}
\bibinfo{author}{\bibfnamefont{H.}~\bibnamefont{Feshbach}},
  \bibinfo{journal}{Ann. Phys.} \textbf{\bibinfo{volume}{19}},
  \bibinfo{pages}{287} (\bibinfo{year}{1962}).

\bibitem[{\citenamefont{Moerdijk et~al.}(1995)\citenamefont{Moerdijk, Verhaar,
  and Axelsson}}]{moerdijk:95}
\bibinfo{author}{\bibfnamefont{A.~J.} \bibnamefont{Moerdijk}},
  \bibinfo{author}{\bibfnamefont{B.~J.} \bibnamefont{Verhaar}},
  \bibnamefont{and} \bibinfo{author}{\bibfnamefont{A.}~\bibnamefont{Axelsson}},
  \bibinfo{journal}{Phys. Rev. A} \textbf{\bibinfo{volume}{51}},
  \bibinfo{pages}{4852} (\bibinfo{year}{1995}).

\bibitem[{\citenamefont{Marcelis et~al.}(2004)\citenamefont{Marcelis, van
  Kempen, Verhaar, and Kokkelmans}}]{marcelis:04}
\bibinfo{author}{\bibfnamefont{B.}~\bibnamefont{Marcelis}},
  \bibinfo{author}{\bibfnamefont{E.~G.~M.} \bibnamefont{van Kempen}},
  \bibinfo{author}{\bibfnamefont{B.~J.} \bibnamefont{Verhaar}},
  \bibnamefont{and} \bibinfo{author}{\bibfnamefont{S.~J.~J.~M.~F.}
  \bibnamefont{Kokkelmans}}, \bibinfo{journal}{Phys. Rev. A}
  \textbf{\bibinfo{volume}{70}}, \bibinfo{pages}{012701}
  (\bibinfo{year}{2004}).

\bibitem[{\citenamefont{Moerdijk et~al.}(1994)\citenamefont{Moerdijk, Stwalley,
  Hulet, and Verhaar}}]{moerdijk:94}
\bibinfo{author}{\bibfnamefont{A.~J.} \bibnamefont{Moerdijk}},
  \bibinfo{author}{\bibfnamefont{W.~C.} \bibnamefont{Stwalley}},
  \bibinfo{author}{\bibfnamefont{R.~G.} \bibnamefont{Hulet}}, \bibnamefont{and}
  \bibinfo{author}{\bibfnamefont{B.~J.} \bibnamefont{Verhaar}},
  \bibinfo{journal}{Phys. Rev. Lett.} \textbf{\bibinfo{volume}{72}},
  \bibinfo{pages}{40} (\bibinfo{year}{1994}).

\bibitem[{\citenamefont{Yan and Babb}(1998)}]{yan:98}
\bibinfo{author}{\bibfnamefont{Z.-C.} \bibnamefont{Yan}} \bibnamefont{and}
  \bibinfo{author}{\bibfnamefont{J.~F.} \bibnamefont{Babb}},
  \bibinfo{journal}{Phys. Rev. A} \textbf{\bibinfo{volume}{58}},
  \bibinfo{pages}{1247} (\bibinfo{year}{1998}).

\bibitem[{\citenamefont{Przybytek and Jeziorski}()}]{przybytek:tobepublished}
\bibinfo{author}{\bibfnamefont{M.}~\bibnamefont{Przybytek}} \bibnamefont{and}
  \bibinfo{author}{\bibfnamefont{B.}~\bibnamefont{Jeziorski}},
  \bibinfo{note}{to be published}.

\bibitem[{\citenamefont{Leo et~al.}(2001)\citenamefont{Leo, Venturi,
  Whittingham, and Babb}}]{leo:01}
\bibinfo{author}{\bibfnamefont{P.~J.} \bibnamefont{Leo}},
  \bibinfo{author}{\bibfnamefont{V.}~\bibnamefont{Venturi}},
  \bibinfo{author}{\bibfnamefont{I.~B.} \bibnamefont{Whittingham}},
  \bibnamefont{and} \bibinfo{author}{\bibfnamefont{J.~F.} \bibnamefont{Babb}},
  \bibinfo{journal}{Phys. Rev. A} \textbf{\bibinfo{volume}{64}},
  \bibinfo{pages}{042710} (\bibinfo{year}{2001}).

\bibitem[{\citenamefont{Tang et~al.}(1998)\citenamefont{Tang, Toennies, and
  Yui}}]{tang:98}
\bibinfo{author}{\bibfnamefont{K.~T.} \bibnamefont{Tang}},
  \bibinfo{author}{\bibfnamefont{J.~P.} \bibnamefont{Toennies}},
  \bibnamefont{and} \bibinfo{author}{\bibfnamefont{C.~L.} \bibnamefont{Yui}},
  \bibinfo{journal}{Int. Rev. Phys. Chem.} \textbf{\bibinfo{volume}{17}},
  \bibinfo{pages}{363} (\bibinfo{year}{1998}).

\bibitem[{\citenamefont{Fedichev et~al.}(1996)\citenamefont{Fedichev, Reynolds,
  Rahmanov, and Shlyapnikov}}]{fedichev:96}
\bibinfo{author}{\bibfnamefont{P.~O.} \bibnamefont{Fedichev}},
  \bibinfo{author}{\bibfnamefont{M.~W.} \bibnamefont{Reynolds}},
  \bibinfo{author}{\bibfnamefont{U.~M.} \bibnamefont{Rahmanov}},
  \bibnamefont{and} \bibinfo{author}{\bibfnamefont{G.~V.}
  \bibnamefont{Shlyapnikov}}, \bibinfo{journal}{Phys. Rev. A}
  \textbf{\bibinfo{volume}{53}}, \bibinfo{pages}{1447} (\bibinfo{year}{1996}).

\bibitem[{\citenamefont{D'Incao and Esry}(2005)}]{incao:05}
\bibinfo{author}{\bibfnamefont{J.~P.} \bibnamefont{D'Incao}} \bibnamefont{and}
  \bibinfo{author}{\bibfnamefont{B.~D.} \bibnamefont{Esry}},
  \bibinfo{journal}{Phys. Rev. Lett.} \textbf{\bibinfo{volume}{94}},
  \bibinfo{pages}{213201} (\bibinfo{year}{2005}).

\bibitem[{\citenamefont{Venturi and Whittingham}(2000)}]{venturi:00}
\bibinfo{author}{\bibfnamefont{V.}~\bibnamefont{Venturi}} \bibnamefont{and}
  \bibinfo{author}{\bibfnamefont{I.~B.} \bibnamefont{Whittingham}},
  \bibinfo{journal}{Phys. Rev. A} \textbf{\bibinfo{volume}{61}},
  \bibinfo{pages}{060703} (\bibinfo{year}{2000}).

\bibitem[{\citenamefont{Stas et~al.}(2006)\citenamefont{Stas, McNamara,
  Hogervorst, and Vassen}}]{stas:06}
\bibinfo{author}{\bibfnamefont{R.~J.~W.} \bibnamefont{Stas}},
  \bibinfo{author}{\bibfnamefont{J.~M.} \bibnamefont{McNamara}},
  \bibinfo{author}{\bibfnamefont{W.}~\bibnamefont{Hogervorst}},
  \bibnamefont{and} \bibinfo{author}{\bibfnamefont{W.}~\bibnamefont{Vassen}},
  \bibinfo{journal}{Phys. Rev. A} \textbf{\bibinfo{volume}{73}},
  \bibinfo{eid}{032713} (\bibinfo{year}{2006}).

\bibitem[{\citenamefont{McNamara et~al.}(2007)\citenamefont{McNamara, Stas,
  Hogervorst, and Vassen}}]{mcnamara:07}
\bibinfo{author}{\bibfnamefont{J.~M.} \bibnamefont{McNamara}},
  \bibinfo{author}{\bibfnamefont{R.~J.~W.} \bibnamefont{Stas}},
  \bibinfo{author}{\bibfnamefont{W.}~\bibnamefont{Hogervorst}},
  \bibnamefont{and} \bibinfo{author}{\bibfnamefont{W.}~\bibnamefont{Vassen}},
  \bibinfo{journal}{Phys. Rev. A} \textbf{\bibinfo{volume}{75}},
  \bibinfo{pages}{062715} (\bibinfo{year}{2007}).

\bibitem[{\citenamefont{Simoni et~al.}(2008)\citenamefont{Simoni, Zaccanti,
  D'Errico, Fattori, Roati, Inguscio, and Modugno}}]{simoni:08}
\bibinfo{author}{\bibfnamefont{A.}~\bibnamefont{Simoni}},
  \bibinfo{author}{\bibfnamefont{M.}~\bibnamefont{Zaccanti}},
  \bibinfo{author}{\bibfnamefont{C.}~\bibnamefont{D'Errico}},
  \bibinfo{author}{\bibfnamefont{M.}~\bibnamefont{Fattori}},
  \bibinfo{author}{\bibfnamefont{G.}~\bibnamefont{Roati}},
  \bibinfo{author}{\bibfnamefont{M.}~\bibnamefont{Inguscio}}, \bibnamefont{and}
  \bibinfo{author}{\bibfnamefont{G.}~\bibnamefont{Modugno}},
  \bibinfo{journal}{Phys. Rev. A} \textbf{\bibinfo{volume}{77}},
  \bibinfo{pages}{052705} (\bibinfo{year}{2008}).

\bibitem[{\citenamefont{van Kempen et~al.}(2002)\citenamefont{van Kempen,
  Kokkelmans, Heinzen, and Verhaar}}]{kempen:02}
\bibinfo{author}{\bibfnamefont{E.~G.~M.} \bibnamefont{van Kempen}},
  \bibinfo{author}{\bibfnamefont{S.~J. J. M.~F.} \bibnamefont{Kokkelmans}},
  \bibinfo{author}{\bibfnamefont{D.~J.} \bibnamefont{Heinzen}},
  \bibnamefont{and} \bibinfo{author}{\bibfnamefont{B.~J.}
  \bibnamefont{Verhaar}}, \bibinfo{journal}{Phys. Rev. Lett.}
  \textbf{\bibinfo{volume}{88}}, \bibinfo{pages}{093201}
  (\bibinfo{year}{2002}).

\bibitem[{\citenamefont{Hutson}(2007)}]{hutson:07}
\bibinfo{author}{\bibfnamefont{J.~M.} \bibnamefont{Hutson}},
  \bibinfo{journal}{New J. Phys.} \textbf{\bibinfo{volume}{9}}
  (\bibinfo{year}{2007}).

\bibitem[{\citenamefont{Smirne et~al.}(2007)\citenamefont{Smirne, Godun,
  Cassettari, Boyer, Foot, Volz, Syassen, D\"urr, Rempe, Lee
  et~al.}}]{smirne07}
\bibinfo{author}{\bibfnamefont{G.}~\bibnamefont{Smirne}},
  \bibinfo{author}{\bibfnamefont{R.~M.} \bibnamefont{Godun}},
  \bibinfo{author}{\bibfnamefont{D.}~\bibnamefont{Cassettari}},
  \bibinfo{author}{\bibfnamefont{V.}~\bibnamefont{Boyer}},
  \bibinfo{author}{\bibfnamefont{C.~J.} \bibnamefont{Foot}},
  \bibinfo{author}{\bibfnamefont{T.}~\bibnamefont{Volz}},
  \bibinfo{author}{\bibfnamefont{N.}~\bibnamefont{Syassen}},
  \bibinfo{author}{\bibfnamefont{S.}~\bibnamefont{D\"urr}},
  \bibinfo{author}{\bibfnamefont{G.}~\bibnamefont{Rempe}},
  \bibinfo{author}{\bibfnamefont{M.~D.} \bibnamefont{Lee}},
  \bibnamefont{et~al.}, \bibinfo{journal}{Phys. Rev. A}
  \textbf{\bibinfo{volume}{75}}, \bibinfo{pages}{020702}
  (\bibinfo{year}{2007}).

\bibitem[{\citenamefont{Hutson et~al.}(2009)\citenamefont{Hutson, Beyene, and
  Gonz\'alez-Mart\'\i{}nez}}]{hutson:09}
\bibinfo{author}{\bibfnamefont{J.~M.} \bibnamefont{Hutson}},
  \bibinfo{author}{\bibfnamefont{M.}~\bibnamefont{Beyene}}, \bibnamefont{and}
  \bibinfo{author}{\bibfnamefont{M.~L.}
  \bibnamefont{Gonz\'alez-Mart\'\i{}nez}}, \bibinfo{journal}{Phys. Rev. Lett.}
  \textbf{\bibinfo{volume}{103}}, \bibinfo{pages}{163201}
  (\bibinfo{year}{2009}).

\bibitem[{\citenamefont{Da\ifmmode~\mbox{\c{}}\else
  \c{}\fi{}browski}(1996)}]{dabrowski:96}
\bibinfo{author}{\bibfnamefont{J.}~\bibnamefont{Da\ifmmode~\mbox{\c{}}\else
  \c{}\fi{}browski}}, \bibinfo{journal}{Phys. Rev. C}
  \textbf{\bibinfo{volume}{53}}, \bibinfo{pages}{2004} (\bibinfo{year}{1996}).

\bibitem[{\citenamefont{Beams et~al.}(2006)\citenamefont{Beams, Peach, and
  Whittingham}}]{beams:06}
\bibinfo{author}{\bibfnamefont{T.~J.} \bibnamefont{Beams}},
  \bibinfo{author}{\bibfnamefont{G.}~\bibnamefont{Peach}}, \bibnamefont{and}
  \bibinfo{author}{\bibfnamefont{I.~B.} \bibnamefont{Whittingham}},
  \bibinfo{journal}{Phys. Rev. A} \textbf{\bibinfo{volume}{74}},
  \bibinfo{pages}{014702} (\bibinfo{year}{2006}).

\bibitem[{\citenamefont{Moal et~al.}(2007)\citenamefont{Moal, Portier, Zahzam,
  and Leduc}}]{moal:07}
\bibinfo{author}{\bibfnamefont{S.}~\bibnamefont{Moal}},
  \bibinfo{author}{\bibfnamefont{M.}~\bibnamefont{Portier}},
  \bibinfo{author}{\bibfnamefont{N.}~\bibnamefont{Zahzam}}, \bibnamefont{and}
  \bibinfo{author}{\bibfnamefont{M.}~\bibnamefont{Leduc}},
  \bibinfo{journal}{Phys. Rev. A} \textbf{\bibinfo{volume}{75}},
  \bibinfo{pages}{033415} (\bibinfo{year}{2007}).

\end{thebibliography}

\end{document}